\documentclass[manuscript]{aastex} 
\usepackage{graphicx,epsfig}
\usepackage{natbib}                              
\usepackage{rotating}
\usepackage{txfonts}
\usepackage{longtable,lscape,url}
\usepackage{rotating}

\shorttitle{Characterization of a sample of intermediate-type AGN I.}
\shortauthors{Ben\'itez et al.}

\begin{document}

\title{CHARACTERIZATION OF A SAMPLE OF INTERMEDIATE-TYPE AGN. I. Spectroscopic properties and serendipitous discovery of new Dual AGN}
\shorttitle{}
\author{Erika Ben\'itez\altaffilmark{1}, Jairo M\'endez-Abreu\altaffilmark{2,3}, Isaura Fuentes-Carrera\altaffilmark{4}, Irene Cruz-Gonz\'alez\altaffilmark{1}, 
Benoni Mart\'inez\altaffilmark{1}, Luis L\'opez-Martin\altaffilmark{2,3}, Elena Jim\'enez-Bail\'on\altaffilmark{1}, Jonathan Le\'on-Tavares\altaffilmark{5}, Vahram H. Chavushyan\altaffilmark{6}}

\altaffiltext{1}{Instituto de Astronom\'ia, Universidad Nacional Aut\'onoma de M\'exico, Apdo. Postal 70-264, M\'exico D.F. 04510, M\'exico}

\altaffiltext{2}{Instituto de Astrof\'{\i}sica de Canarias, 38200 La Laguna, Tenerife, Spain}

\altaffiltext{3}{Departamento de Astrof\'isica, Universidad de La Laguna, E-38205 La Laguna, Tenerife, Spain}

\altaffiltext{4}{Escuela Superior de F\'isica y Matem\'aticas, Instituto Polit\'ecnico Nacional (ESFM-IPN), U.P. Adolfo L\'opez Mateos, M\'exico D.F. 07730, M\'exico}

\altaffiltext{5}{Aalto University Mets\"ahovi Radio Observatory, Mets\"ahovintie 114, 02540, Kylm\"al\"a, Finland}

\altaffiltext{6}{Instituto Nacional de Astrof\'{\i}sica, \'Optica y Electr\'onica, Apdo. Postal 51-216, 72000 Puebla, M\'exico}

\email{erika@astro.unam.mx}

\accepted{Nov 26 2012}

\begin{abstract}

A sample of 10 nearby intermediate-type active galactic nuclei (AGN) drawn from the Sloan Digital Sky Survey 
(SDSS-DR7) is presented.   The aim of this work is to provide estimations of the 
black hole mass for the sample galaxies from the dynamics of the broad line region. For this purpose,
a detailed spectroscopic analysis of the objects was done. Using BPT diagnostic diagrams we have carefully 
classified the objects as true intermediate-type AGN and found that 80\%$^{+7.2\%}_{-17.3\%}$ are composite AGN. 
The black hole mass estimated for the sample is within  6.54$\pm$0.16\,$<$\,log\,$M_{\rm BH}$\,$<$\,7.81$\pm$0.14.
Profile analysis show that five objects (\object{J120655.63+501737.1}, 
\object{J121607.08+504930.0}, \object{J141238.14+391836.5}, \object{J143031.18+524225.8} and 
\object{J162952.88+242638.3}) have narrow double-peaked emission lines in both the red 
(H$\alpha$, [\ion{N}{2}]$\lambda\lambda$6548,6583 and 
[\ion{S}{2}]$\lambda\lambda$6716,6731) and the blue  (H$\beta$ and 
[\ion{O}{3}]$\lambda\lambda$4959,5007) region of the spectra, 
with velocity differences ($\Delta V$) between the double peaks within 114\,$<\Delta V\,<$\,256 km~s$^{-1}$. Two of them,
\object{J121607.08+504930.0} and \object{J141238.14+391836.5} are candidates for dual AGN since their double-peaked emission 
lines are dominated by AGN activity. In searches of dual AGN; Type 1, Type 1I and intermediate-type AGN should 
be carefully separated, due to the high serendipitous number of narrow double-peaked sources
(50\%$\pm$14.4\%) found in our sample.

\end{abstract}





\section{Introduction}\label{introduction}

Among nearby active galactic nuclei (AGN) the so-called Seyfert (Sy) galaxies are commonly found. Seyfert galaxies
properties were described in the early work by \citet{1974ApJ...192..581K} who classified them as Sy 1 and Sy 2 
depending on the line widths observed in their optical spectra.  A few years later, \citet{1981ApJ...249..462O}
introduced different sub-classes of Sy galaxies according to  the relative intensity showed by the broad emission lines in their spectra. 
In particular, he introduced the intermediate Sy class, spanning from Sy 1.2  to Sy 1.9. For instance, a Sy 1.8 galaxy shows weak, 
but readily visible broad H$\alpha$ and H$\beta$ emission lines, while a  Sy 1.9 shows only a broad H$\alpha$  component. 
Recently, \citet{2008NewAR..52..227T} reviewed the general classification and unification of AGN where Sy 1 
and Sy 2 galaxies are more generally grouped as Type 1 and Type 2 AGN. Since intermediate-type Sy galaxies do show broad emission 
line components, they belong to the Type 1 class, but they are more difficult to find since sometimes the intensity of these components 
can be very low. We will refer to these objects as intermediate-type AGN to distinguish them from the more general Type 1 class.

Early studies on intermediate-type AGN, have proposed that the origin of spectroscopically observed weak emission lines
produced in the clouds of the Broad Line Region (BLR) could be explained by internal reddening 
\citep{1981ApJ...249..462O,1990ApJ...355...88G, 1995ApJ...440..141G}.
Another explanation based in the unified model for AGN \citep{1995PASP..107..803U} assumes that a dusty-torus located some 
parsecs away from the central source blocks part of the emission. Other possibility is that their weak lines could be due to an intrinsically faint non-thermal continuum 
emission, maybe due to variations in the ionizing continuum \citep[see][and references therein]{2010ApJ...725.1749T}.

On the other hand, the availability of high-quality spectra of AGN in the Sloan Digital Sky Survey \citep[SDSS;][]{2000AJ....120.1579Y} 
makes it possible to identify intermediate-type AGN and furthermore study them as a class, in order to find their place in the
black hole demography for nearby AGN \citep[see][]{2004cBHg.symp..292H}. In general, new estimates of fundamental parameters 
like the black hole (BH) mass or M$_{BH}$ in AGN can contribute to the understanding of the associated fueling 
mechanisms. The M$_{BH}$ can be estimated using different empirically calibrated scaling relations 
in samples designed to study particular types of AGN \citep[e.g.][]{1999ApJ...526..579W,2006ApJ...641..689V}. In particular, 
BH mass estimates can be obtained using the correlation between the AGN continuum luminosity 
\citep{2000ApJ...533..631K, 2005ApJ...629...61K} and BLR radius. By combining continuum luminosity with the width of the H$\beta$ emission line (in objects that have z$<$1, otherwise this line goes to the IR) it is possible to estimate the mass of the black hole by using a single AGN spectrum, yielding
BH masses that are accurate to a factor of $\sim$3 \citep[][]{2006ApJ...641..689V}.

In the case of intermediate-type AGN, estimates of the M$_{BH}$ need to be done systematically. Previous works in this type
of AGN are based on single or very few objects \citep[e.g.][]{1981ApJ...249..462O, 1989ApJ...340..190G, 1993A&A...275..451R, 1996MNRAS.280....6X, 
1997POBeo..57...95P, 2000ApJ...532L..17Q, 2006RMxAA..42....3T}. A larger amount of objects was recently studied by \citet{2010ApJ...725.1749T}. 
These authors selected a sample of 34 previously classified Sy1.8/1.9 and found that 18 are missclassified objects.
Therefore, it is necessary to carefully classify candidates in order to establish if they conform a true sample of intermediate-type AGN.

Among the Type 1 and Type 2 AGN samples drawn from the SDSS, the sample of double-peaked narrow emission line AGN stands out.
Usually they show {\textsc{[O\,III]} line profiles with velocities splitting in the range 151 to 1314 km\,s$^{-1}$.  
The relevance of studying these kind of AGN is based on the idea that they are candidates for sub-kpc or kpc-scale dual 
AGN \citep[e.g.][]{2009ApJ...702L..82C}. Discovering new dual (or binary, if their nuclei appears separated on pc-scales) AGN candidates 
is fundamental to constrain models of galaxy formation and evolution. Since it is commonly believed that mergers can trigger or enhance 
nuclear accretion, it is fundamental to establish their frequency by means of detecting dual or binary AGN.  
Nevertheless, it is a fact that direct observational evidence for binary super massive black holes still remains scarce 
\citep[see][and references therein]{2011ApJ...738L...2M}. 

Double-peaked {\textsc{[O\,III]} line profiles could arise from several mechanisms \citep[e.g.,][]{2009ApJ...705L..20X,2010ApJ...716..131R,2011ApJ...733..103F,2011ApJ...735...48S}: 
(1) the orbital motion of a binary AGN,  (2) gas kinematics in extended narrow-line regions, and (3) unresolved nuclear gas kinematics 
(e.g., aligned outflows or disk rotation on small scales). \citet{2011arXiv1107.3564F} have obtained high-resolution imaging of 106 narrow 
double-peaked AGN (37 Type 1 and 69 Type 1I) and found that 31 have companions within 3 \arcsec.
Their integral field spectroscopic observations has allowed them to group the narrow double-peaked AGN in the above three broad categories, 
according to the origin of the double-peaked {\textsc{[O\,III]} line profiles, finding that only 4.5-12 \% are binaries. Other studies on the frequency of these objects among AGN 
\citep[e.g.,][]{2009ApJ...705L..76W,2010ApJ...708..427L, 2010ApJ...716..866S} have show that
there are 340 objects between 0.008 $<\,z\,<$0.686 or, equivalently, $\sim$1 \% of the entire SDSS AGN sample harboring possible dual AGN. 
Nevertheless, a recent study done by \citet{2012ApJ...745...67F} shows that several objects classified in the literature as ``binary'' are, 
most likely, single AGN with extended narrow-line regions. Therefore, the frequency of double-peak objects and binaries among intermediate-type AGN are still unknown, and so this is worth to investigate.

The aim of this work is to isolate inter\-mediate-type AGN objects and estimate their black hole mass.
In particular, we are interested in providing new estimates that can contribute to the 
study of the black hole demography in nearby AGN.
For this purpose, we present the spectral analysis of a sample of 10 nearby ($10^{42}\,\lesssim\,L_{\rm bol}\,\lesssim\,10^{44}$erg\,s$^{-1}$)
intermediate-type AGN (i.e. intermediate Sy galaxies) . Spectral analysis will provide a more precise classification through the use of the 
Baldwin-Phillips-Terlevich (BPT) diagnostic diagrams. A detailed profile decomposition allowed us to discover  
new narrow double-peaked AGN candidates among the objects in the sample.

The paper  is organized  as follows: In  Sect.~\ref{sample} we describe the sample selection and in Sect.~\ref{analysis}
the analysis of the SDSS spectroscopic data.  Sect.~\ref{results} shows  the  results obtained for the sample and 
relevant properties of some individual objects. The discussion and conclusions are presented in Sect.~\ref{discussion}. 
An Appendix on the available X-ray data for one of the selected sources is presented. 
The cosmology adopted for this work  is  $H_{0}=$70  km  s$^{-1}$  Mpc$^{-1}$,  $\Omega_{m}=$0.3  and $\Omega_{\lambda}=$0.7.

\section{Sample Selection}
\label{sample}

We selected objects flagged  as quasars (QSO) in the SDSS database that have Petrosian $g$-band magnitudes
between 14 $\lesssim\,m_g\,\lesssim$ 17 and redshifts within 0.026 $<z <$ 0.12, i.e., a sample of nearby AGN.  From this sample, we chose 
only those galaxies that have spectroscopic data in the SDSS  database in order to accurately obtain their spectral classification.
An additional condition imposed was that visually all selected galaxies should have spectral characteristics of intermediate-type AGN. 
We ended up with a sample of 23 candidates that were photometrically observed with the Nordic Optical Telescope (NOT). Results obtained with
these data are presented in a companion paper (Ben\'itez et al. 2012b, hereafter Paper II). 
A final sample of 10 candidates for intermediate-type AGN were chosen for this work, these are presented in Table~\ref{tab:sample}. 
The remaining objects were taken out from the sample since after a preliminary 
analysis of their spectra, we found that some of them were either narrow line Seyfert 1 \citep{1985ApJ...297..166O} or Seyfert 1 objects.  

\section{Analysis of SDSS-DR7 spectra}
\label{analysis}

The optical SDSS-DR7 spectra of our sample were processed using the STARLIGHT code \citep{2005MNRAS.358..363C,2006MNRAS.370..721M}. 
This code has been successfully used to study some AGN samples \citep[see][]{2007ApJ...668..721B,2006MNRAS.371..972S,2009NewAR..53..186R,2011MNRAS.413.1687C,2011RMxAA..47..361C,2011A&A...527A..38M,2011MNRAS.411.1127L}. 
In a first run, all spectra were corrected for Galactic extinction using the dust maps of \citet{1998ApJ...500..525S} and the extinction law of \citet{1989ApJ...345..245C}.  Also in this first run, in order to obtain the signal-to-noise ratio (S/N) we have masked the most prominent emission lines that are known to be present in the composite spectra of QSOs \citep[see][]{1991ApJ...373..465F}. So, the following lines were masked: [\ion{O}{2}]$\lambda\lambda$3726,3729, [\ion{Ne}{3}]$\lambda$3869, H$\epsilon$\,$\lambda$3970, H$\delta$\,$\lambda$4102, H$\gamma$\,$\lambda$4340, H$\beta$\,$\lambda$4861, [\ion{O}{3}]$\lambda\lambda$4959,5007, \ion{He}{1}$\lambda$5876, [\ion{O}{1}]$\lambda$6300, H$\alpha$\,$\lambda$6562, [\ion{N}{2}]$\lambda\lambda$6548,6583, and [\ion{S}{2}]$\lambda\lambda$6717,6731; together with the ISM absorption from Na\,D$\lambda$5890 \AA). The signal-to-noise ratio (S/N) was obtained for each spectrum in the wavelength range of 4730 to 4780 \AA~  and is presented in Col.~2 of Table~\ref{tab:fits}. 

Based on the S/N previously obtained, each spectrum was randomly perturbed 100 times using a Monte Carlo (MC) seed. Our lowest S/N value was 5.9, so the perturbations varied $\pm$16\% of the signal value. We obtained 100 simulated spectra per object that were also analyzed with STARLIGHT, masking again the prominent emission lines as we did in the first run. From these simulated spectra, we obtained a collection of 100 models for the absorption spectrum of the host galaxy, from which we will obtain an estimation of the velocity dispersion. We chose the median of the distribution of models as our best model of the host-galaxy absorption spectrum. The stellar velocity dispersion $\sigma_{\star}$ was estimated from the median value of the distribution and the quoted errors are the rms values (see Table \ref{tab:BHm}). In all cases the obtained $\sigma_{\star}$ is larger than 65\,km~s$^{-1}$, which is the nominal SDSS velocity resolution \citep[e.g.,][]{2010ApJ...708..427L}. The AGN continuum flux at 5100 \AA\, was estimated from the the collection of 100 power-law fittings. Again we chose the best fit from the median value of the distribution of models to obtain F 5100 \AA\ and the rms as the associated error.
This flux will be used to derive the BH mass, as is described in  Sect.~\ref{virial}. Therefore, STARLIGHT provided us with the best synthetic spectra for our AGN sample. 

The original SDSS spectra and the final pure AGN spectra, together with the obtained fits for the sample objects are shown in Figs.~\ref{fig:stl1} to \ref{fig:stl9}. 
These figures show how the stellar contribution is subtracted with STARLIGHT from the composite spectra. This is essential for our study since the absorption synthetic spectra were used to estimate the velocity dispersion and the pure AGN spectra to perform a line profile analysis. An example of the best model obtained for object \#9 is shown in Figure \ref{fig:best}, where it can be seen how STARLIGHT is capable to fit the absorption lines used to estimate the velocity dispersion along with the stellar continuum. The absorption lines used to estimate $\sigma_{\star}$ are marked in the blow-up spectrum.  

The complete optical spectroscopic parameters obtained for the sample are presented in Table~\ref{tab:fits}.  
The FWHM for both the H$\alpha$ and H$\beta$ broad components have errors $<$10\%. The line
ratios have also errors $<$10\% except for object \#10 that has errors $\sim$18\%.
The profile decomposition in all cases was done using the algorithm  PeakFit \footnote{Systat Software Inc. \url{http://www.sigmaplot.com}}. 
From our profile decomposition analysis, we find that in five objects (\#1, \#3, \#4, \#5 and \#8) the
[\ion{O}{3}]$\lambda\lambda$4959,5007, [\ion{S}{2}]$\lambda\lambda$6716,6731, [\ion{N}{2}] $\lambda\lambda$6548,6583 doublets
and the narrow components of H$\alpha$ and H$\beta$ clearly show a double-peaked profile. The spectral profile decomposition of these five narrow-double 
peaked AGN are shown in Figs. \ref{fig:dblpk1}  to  \ref{fig:dblpk8}. We  have checked the redshifts using some absorption lines from the host galaxy spectra 
and found that they are in good agreement with the values reported in NED\footnote{The NASA/IPAC Extragalactic Database (NED) is operated by the Jet  Propulsion Laboratory, California  Institute of Technology, under   contract   with   the   National   Aeronautics   and   Space  Administration.}, which are based mainly on emission lines. Therefore, the Gaussians centroids 
were forced to correspond to the host galaxy redshift. For all objects, the widths of the Gaussians fitted to the narrow lines have the same Doppler broadening, and the intensity ratio
[\ion{O}{3}]$\lambda$5007/[\ion{O}{3}]$\lambda$4959 was fixed to its theoretical value. Table~\ref{tab:deltav} presents the velocity difference defined as $\Delta V\,=\,V_{\rm red}  - V_{\rm blue}$
for each object with double-peaked narrow emission lines for various lines.

\citet{1981PASP...93....5B}  propose a suite of diagnostic diagrams (known as BPT) to classify emission-line galaxies in order to disentangle the dominant energy source. The diagrams are based on four optical line ratios [\ion{O}{3}]/H$\beta$, [\ion{N}{2}]/H$\alpha$, [\ion{S}{2}]/H$\alpha$ and [\ion{O}{1}]/H$\alpha$.  Following \citet{2006MNRAS.372..961K}, we present in  Figure~\ref{fig:diagnostic} two BPT diagnostic diagrams that show the location of the sample objects. These diagrams used  line ratios [\ion{O}{3}]/H$\beta$, [\ion{N}{2}]/H$\alpha$ and [\ion{S}{2}]/H$\alpha$.

We note that for the narrow double-peaked objects we present the obtained line ratios separately using the blue and red components. 
This allowed us to see how objects can change their spectral classification in the diagnostic diagram. We have found that 8 objects are
composite AGN (Starburst plus AGN). In Table \ref{tab:BHm} we present the spectral classification obtained for all objects. In some objects
we obtained different classifications since they appear in different places in the two BPT diagrams. In particular, object \#2 appears as a  
Starburst (SB) galaxy in one of the BPT diagrams. In addition, since all objects present a broad line component, and following 
\citet{1992ApJS...79...49W}, we have obtained their Sy type based on F$_{5007}$/F$_{H\beta}=$\,R. Then, we have assigned Sy 1
(R\,$\leq$\,0.3), Sy 2 (0.3\,$<$\,R\,$\leq$\,1), Sy 1.5 (1\,$<$\,R\,$\leq$\,4), Sy 1.8 (R\,$>$\,4), and Sy 1.9 if only broad H$\alpha$ was seen.

\section{Results}\label{results}

\subsection{$M_{\rm BH}$ estimates}
\label{virial}

The $M_{\rm BH}$ for all objects in the sample were indirectly derived using an empirically calibrated photoionization method \citep{1999ApJ...526..579W, 2002ApJ...571..733V, 2006ApJ...641..689V}. We derived BH masses using the  SDSS-DR7 spectra using equation 5 from \citet{2006ApJ...641..689V}, which makes use of the FWHM of the H$\beta$ emission-line and the continuum luminosity at 5100~\AA. 
Luminosity at 5100 \AA\, ($L_{\rm 5100}$) was estimated from the median flux derived with the power-law fit done with STARLIGHT and the luminosity distances of the objects using the redshifts given in NED.
The bolometric luminosity was obtained using $L_{\rm bol}\sim$9$\lambda$$L_{\lambda5100}$ \citep[see][]{2000ApJ...533..631K}, which yields
the Eddington ratio $L_{\rm bol}/L_{\rm Edd}$, where $L_{\rm Edd}$\-$\approx$1.3$\times$10$^{38}\left(M_{\rm BH}/M_{\odot}\right)$. Our luminosity
results are shown in Table~\ref{tab:lumin}. For objects without H$\beta$ broad  component, the H$\alpha$ line FWHM  was converted to H$\beta$ FWHM using equation 3 given by \citet{2008AJ....135..928S}. 
The BH mass estimates for our sample are shown in Table~\ref{tab:BHm}.

We can also derive the BH mass using the so called M-$\sigma_{\star}$ relation \citep{2000ApJ...539L...9F,2000ApJ...539L..13G}
and see if these estimates are in agreement with the black hole mass derived for these objects using scaling relations. This is the reason why we obtained the velocity dispersion ($\sigma_{\star}$) with STARLIGHT. However, since the use of this relation depends on the bulge type associated with the host galaxy
\citep[e.g.][]{2008MNRAS.386.2242H}, we will present these estimates in Paper II, along with a detailed discussion of the bulge type present in each galaxy based on the photometric and spectroscopic observations.

In Fig.~\ref{fig:mass-lum} we show the BH mass vs. luminosity relation obtained for the objects in our sample.
The diagonal lines show the Eddington limits shown by our objects. Note that internal extinction has not been taken into account. 
All objects have low to moderate accretion Eddington rates ($L_{\rm bol}/L_{\rm Edd}\,<\,$0.11).

\subsection{Double-Peaked Objects}

Spectral analysis showed that among our objects we have five objects that clearly show narrow double-peaked emission lines: 
galaxies \#1, \#3, \#4, \#5 and \#8 (c.f. Table~\ref{tab:sample}). In Table~\ref{tab:deltav} we present the velocity differences, $\Delta
V \,=\,V_{\rm red}  \,-\, V_{\rm blue}$, for H$\alpha$,\,[\ion{N}{2}]$\lambda$6548,\,[\ion{N}{2}]$\lambda$6583,\,[\ion{S}{2}]$\lambda$6716,\,  
[\ion{S}{2}]$\lambda$6731, H$\beta$,\,[\ion{O}{3}]$\lambda$4959, and [\ion{O}{3}]$\lambda$5007 lines. Our definition of a narrow line double-peaked
object requires that in all these lines a double-peaked Gaussian components should be present. The average values and standard deviation 
obtained for each source are listed in the bottom line of this table, which shows that they are within 114\,$\lesssim\,\Delta V\,\lesssim$ 256 km~s$^{-1}$.

\subsection{Notes on particular objects}
\label{individual}

\subsubsection{{J120655.63+501737.1 (\#1)}}

The profile decomposition shows that this AGN was best fitted using double-peaked lines for its narrow line region (NLR,  see Fig.~\ref{fig:dblpk1}). 
Its spectrum was fitted with a broad component for the H$\beta$ and H$\alpha$ lines, and from its locus in the BPT diagrams shown in Fig.~\ref{fig:diagnostic}, 
we classify this object as a composite AGN. The blue component indicates that it is a Sy 1.8+SB, and the red component that it is a Sy 1.5\,+\,SB.

\subsubsection{{J121600.04+124114.3 (\#2)}}

In \citet{2010A&A...518A..10V} and in NED this object is classified as a Sy 1.9, and  in \citet{1986ApJS...62..751M} it is classified as a Starburst (SB) galaxy. 
We agree with both classifications since this object appears in one diagram as a composite Sy 1.9\,+\,SB, and in the other as a clear Starburst galaxy, 
c.f., Fig.~\ref{fig:diagnostic}. We noted that the emission line fitting of this object clearly needed a double-peaked Gaussian component for [\ion{O}{3}], [\ion{S}{2}] and [\ion{N}{2}] lines; but when fitting the narrow  components of H$\alpha$ and H$\beta$ only one Gaussian component was required. 
Since we did not needed double-peaked Gaussians to fit the narrow line components of H$\alpha$ and H$\beta$, this object was not classified as double-peaked. 
 
\subsubsection{{J121607.08+504930.0 (\#3)}}

The profile decomposition analysis shows 
that this object is a narrow-line double-peaked AGN, see Fig.~\ref{fig:dblpk3}. Its spectrum was fitted with a broad component for the H$\beta$ and H$\alpha$ lines, 
and from the locus of each component in the BPT diagnostic diagrams shown in Fig.~\ref{fig:diagnostic}, we classify this object as a Sy 1.8 galaxy.

\subsubsection{{J141238.14+391836.5 (\#4)}}

From our profile decomposition, we find that this AGN is a narrow-line double-peaked (Fig.~\ref{fig:dblpk4}) source. In addition, we noted that this object lacks a 
broad component in the H$\beta$ region.  From its locus in the BPT diagrams  in Fig.~\ref{fig:diagnostic}, and using the blue and the red components, we classify this object as a Sy 1.9 galaxy. 
  
\subsubsection{{J143031.18+524225.8 (\#5)}}

This object is also a narrow-line double-peaked source, see Fig.~\ref{fig:dblpk5}. Since we need a broad Gaussian component to fit the H$\alpha$ region, and from
the BPT diagnostic diagrams in Fig.~\ref{fig:diagnostic}, using both the blue and red components we classify this object as a Sy 1.5\,+\,SB.

\subsubsection{{J162952.88+242638.3 (\#8)}}

The [\ion{O}{3}]$\lambda\lambda$4949,5007 doublets clearly show a double-peaked profile and a rather broad wing. We fitted the profile with two sets of 
three Gaussians.  Two Gaussian profiles were used for fitting the narrow component of H$\beta$ (Fig.~\ref{fig:dblpk8}). The profile decomposition shows a broad 
Gaussian component for H$\alpha$. The BPT diagnostic diagrams in Fig.~\ref{fig:diagnostic} clearly show that, using both red and blue components, the object 
is a composite AGN and we classify it as a Sy 1.9\,+\,SB. The results on the spectral analysis of the XMM-Newton data on this source (see Appendix~\ref{appendix}) 
are in agreement with the optical data ones. This is based on the obtained value of the $\Gamma$ index and the low values found for the equivalent $N_{H}$. 
Both values are compatible with Type~I objects \citep[e.g.,][]{2005A&A...432...15P}.
 
\section{Discussion and Conclusions}
\label{discussion}

The sample presented in this work consists of ten nearby intermediate-type AGN, some of them showing a weak broad H$\alpha$ line component.
An example of this was presented in the profile analysis of \object{J162952.88+242638.3}. It clearly shows a rather weak broad component in H$\alpha$ 
(cf. Fig.~\ref{fig:dblpk8}), typical of a Sy 1.9 galaxy. This result is also supported with our analysis of the XMM public available data on this object that showed 
that the power index and the hydrogen column absorption are compatible with Type 1 (c.f., Appendix A) objects. Thus, our Sy 1.9 classification for this object resulted to be in 
disagreement with \citet{2010ApJ...725.1749T}, that recently classify this object as a Seyfert 2 using the same optical and XMM data.

The BPT diagnostic diagram [\ion{O}{3}]/H$\beta$ vs. [\ion{S}{2}]/H$\alpha$ clearly shows that all single-peaked objects are composite AGN.
In the case of the narrow line double-peaked objects detected in this work, three are composite AGN (\#1, \#5, and \#8) and only two of them (\#3 and \#4) are
pure intermediate Sy galaxies. Therefore, 80\%$^{+7.2\%}_{-17.3\%}$ of our intermediate-type objects  
are composite AGN.  It is interesting to note that using the \citet{1992ApJS...79...49W} classification for intermediate Sy galaxies, we confirm that all objects
studied in this work are intermediate-type AGN. 

The FWHM of H$\beta$ component was used to estimate the BH mass in all objects for the first time. We found that the sample studied in this work 
has a BH mass range of 6.54$\pm$0.16\,$<$\,log\,$M_{\rm BH}$\,$< \- $\,7.81$\pm$0.14. In particular three objects (\#5, \#9, and \#10) 
have $M_{\rm BH}$\,$\sim$10$^{6}M_{\odot}$. Recently, \citet{2011Natur.469..374K} 
suggested that small black holes in bulgeless galaxies or pseudobulges grow as low-level Sy. This idea will be explored in
our companion Paper II, where we study the bulge properties of this sample. Also, in Paper II the  M$_{\rm BH}$ estimates obtained in this work 
will be compared with estimates for the same sample through the use of the M-$\sigma_{\star}$ correlation. 
We also found that objects in our sample have low to moderate accretion rates that go from (0.21$\pm$0.23)$\times$10$^{-2}$ to 
(3.44$\pm$0.24)$\times$10$^{-2}$ and are located at the low-tail of bolometric AGN luminosities. 

Among the intermediate-type AGN identified in this sample, 50\%$\pm$14.4\% 
show narrow double-peaked emission lines. From the BPT diagrams and following \citet{2009ApJ...702L..82C} 
we find that two objects \#3 and \#4 are good candidates 
for dual AGN since both the blue and red emission line components are dominated exclusively by AGN activity. 
The other three narrow-double peaked objects \#1, \#5 and \#8 show one emission line component in the AGN region while the other appears 
in the composite  AGN\,+\,SB region. Therefore, the precise classification of each component in the case of double-peaked AGN is essential, 
since it gives hints on the origin of the observed narrow double-peaked lines. This information could be useful in identifying new
dual AGN.

The objects showing narrow double-peaked emission lines found in this work need to be studied through observations enabling spatially resolved imaging and 
spatially resolved spectroscopy. In searches of dual AGN, Type 1, Type 2 and intermediate-type AGN should be carefully separated, due to the high 
serendipitously number of narrow double-peaked sources found in our sample. These objects could be useful for establishing the properties and origin 
of narrow double-peaked emission lines profile in cases where the peak velocity separation lies below $300$\,km\,s$^{-1}$.

\acknowledgements

We want to thank the anonymous referee for useful suggestions and constructive criticism. We also want to thank 
Dr. D. Clark who carefully read the final version of the manuscript. 
EB and BM acknowledges  financial support  from  UNAM-DGAPA-PAPIIT through grant IN116211. JMA is partially funded  by the
Spanish  MICINN under the  Consolider-Ingenio 2010 Program grant CSD2006-00070 and also by the grants AYA2007-67965-C03-01 
and AYA2010-21887-C04-04. IFC thanks the financial support from CONACYT grant 0133520 and IPN-SIP grant 20121700. 
Funding for the Sloan Digital Sky Survey (SDSS) and SDSS-II has been provided by the Alfred P. Sloan Foundation, the Participating Institutions, 
the National Science Foundation, the U.S. Department of Energy, the National Aeronautics and Space Administration, the Japanese 
Monbukagakusho, and the Max Planck Society, and the Higher Education Funding Council for England. The SDSS Web site is \url{http://www.sdss.org/}. 
This research has made use of the NASA/IPAC Extragalactic Database (NED) which is operated by the Jet Propulsion Laboratory, California Institute of 
Technology, under contract with the National Aeronautics and Space Administration.

\bibliography{syintermedias.bib}

\appendix
\section{XMM DATA ANALYSIS OF J162952.88+242638.3 (\#8)}\label{appendix}

We searched for X-ray data available for our  sample objects in Chandra  and XMM-Newton  databases.  We found  that only one of our 
objects has been observed with XMM-Newton. This object  is J162952.88+242638.3, also known as Mrk~883. 
The  three XMM-Newton observations analyzed for this  object were performed 
during August 13th, 15th and 21st 2006 (Obs. ID  0302260101; 0302260701 and 0302261001, respectively). The exposure times  after  flare removal (see below) 
were 7.9, 10.5 and  10.5 ks. All  the observations were processed  using the standard  Science Analysis  System, SAS  v10.3.0 \citep{2004ASPC..314..759G} and using 
the most updated calibration files available in January 2011. The EPIC-pn events were filtered due to high background events using the method described in 
\citet{2004MNRAS.351..161P}. No signs of pile-up or out-of-time events were detected in any of the observations.

The spectra  of each observation were extracted  from circles centered in  the maximum  emission  pixel  and with  radii  which maximized  the
signal-to-noise, i.e.,  22.5 \arcsec. The  background circular regions used for  the spectral analysis were  located close to  the source and
free  of  any  contaminating  source.   Both  response  matrices  were generated  for each observation  using {\tt  arfgen} and  {\tt remgen}
SAS (\url{http://xmm.esac.esa.int/sas/}) tasks.  The spectra were  binned to  have at least 20 counts per bin  in order  to be  able to  use  the modified
$\chi_{\nu}^2$  technique \citep{1977ats..book.....K}
for the  data fit.   All the three spectra were analyzed simultaneously. The shape of the continuum
is  quite  similar  among  the  three observations,  with  a  slight 
increment  of the emission  for the  third observation.   After taking
into     account    the    absorption     due    to     our    Galaxy,
$N_{H}$=$3.83\times10^{20}$cm$^{-2}$,  the continuum  emission  can be
satisfactorily fitted  with a  single absorbed power-law with  a photon
index $\Gamma$=1.7$\pm$0.03  and an  equivalent hydrogen column density  of the
order of $1\times10^{21}$cm$^{-2}$.  Table~\ref{tab:xray} summarizes
the values of the parameters for the best fit model.

None of the observations of this object present an iron line statistically significant, although both a neutral iron line with the energy fixed to 6.4~keV and an iron line with 
the energy left free to vary were tested. During the eight days in which the observations span, moderate flux variability between the two first observations
and the last one was detected.  Neither power-law index nor intrinsic absorption variation were measured. Figure~\ref{fig:xray} summarizes  
the results for the X-ray analysis of SDSS J162953.88+242638.3: the best-fit model for the three spectra with the residuals to the data, and the contour plots for the confidence 
level between the $\Gamma$ index  and the equivalent $N_{H}$ for the three observations.

\begin{figure*}
\includegraphics[angle=0,width=7cm,height=7.5cm]{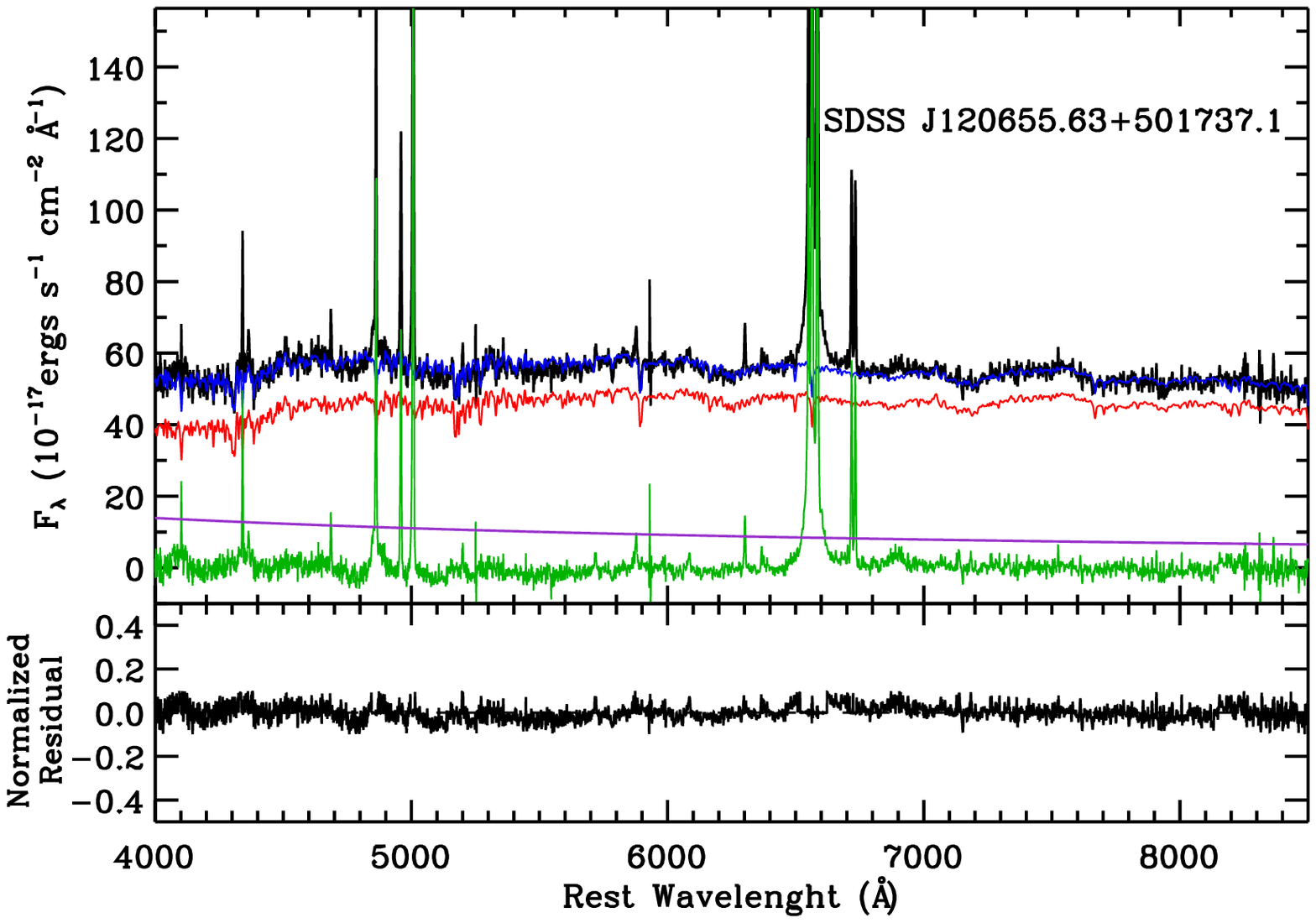}\hspace*{\columnsep}
\includegraphics[angle=0,width=7cm,height=7.5cm]{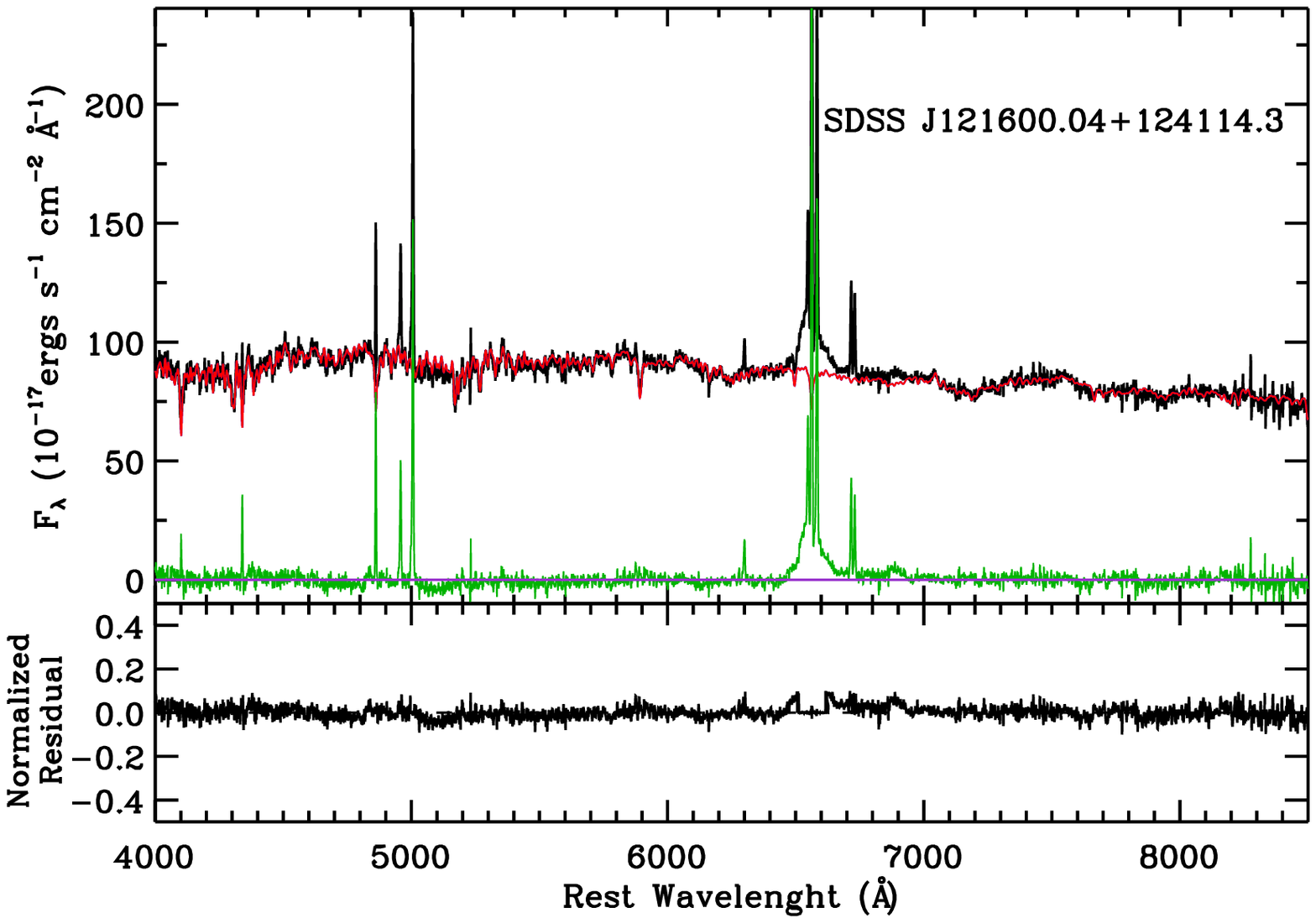}
\caption{STARLIGHT fit of the SDSS spectra. The black line shows the original SDSS de-redshifted spectrum. The red line shows the pure stellar continuum 
spectrum and the purple line the power-law used to fit the AGN continuum. The blue line shows the best fit to the data and represents the sum of the red spectrum plus the AGN continuum. We show in green the final pure AGN spectra of  J120655.63+501737.1 and J121600.04+1241143. Bottom panels show the normalized residuals.}
\label{fig:stl1}
\end{figure*}

\begin{figure*}
\includegraphics[angle=0,width=7cm,height=7.5cm]{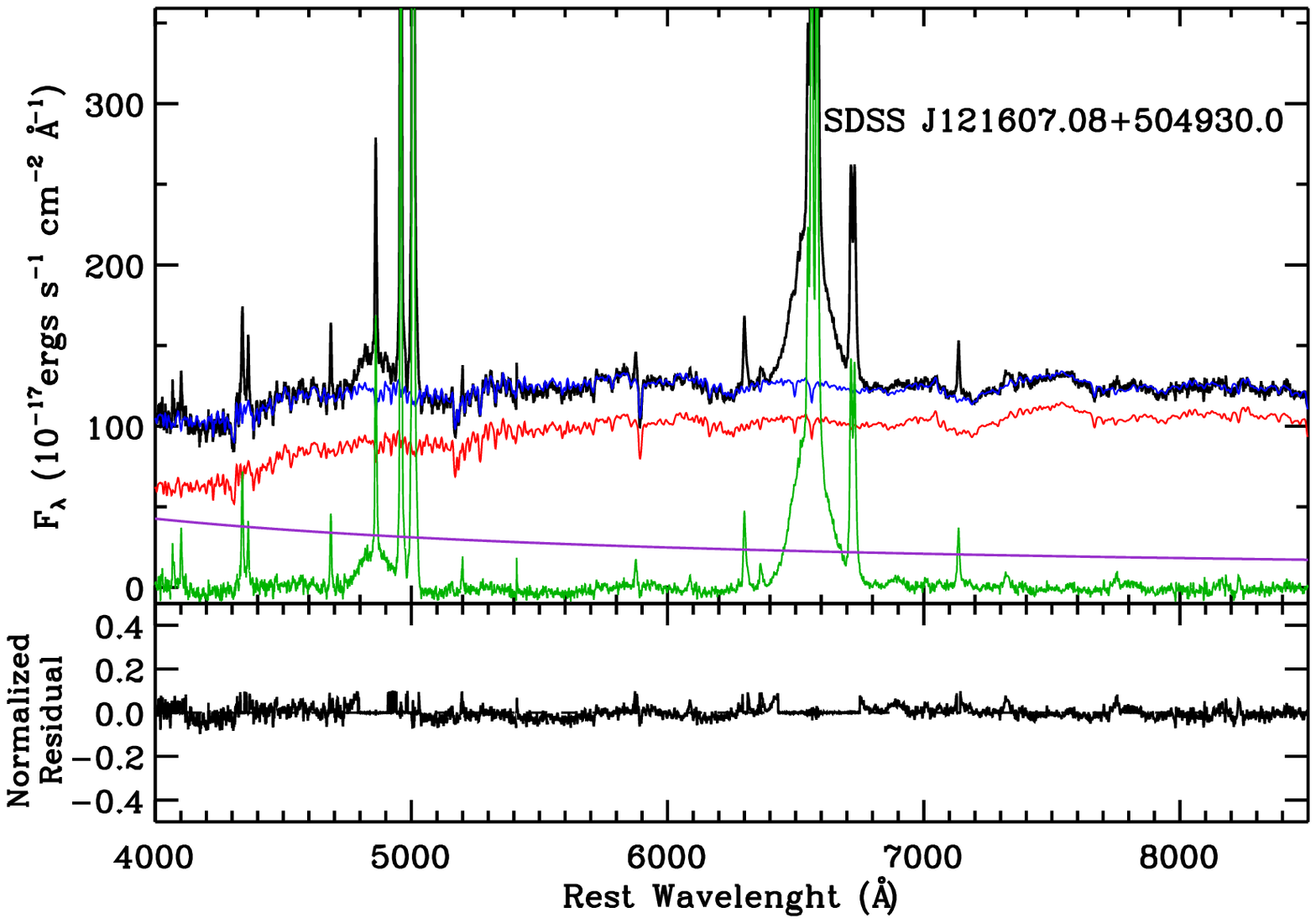}\hspace*{\columnsep}
\includegraphics[angle=0,width=7cm,height=7.5cm]{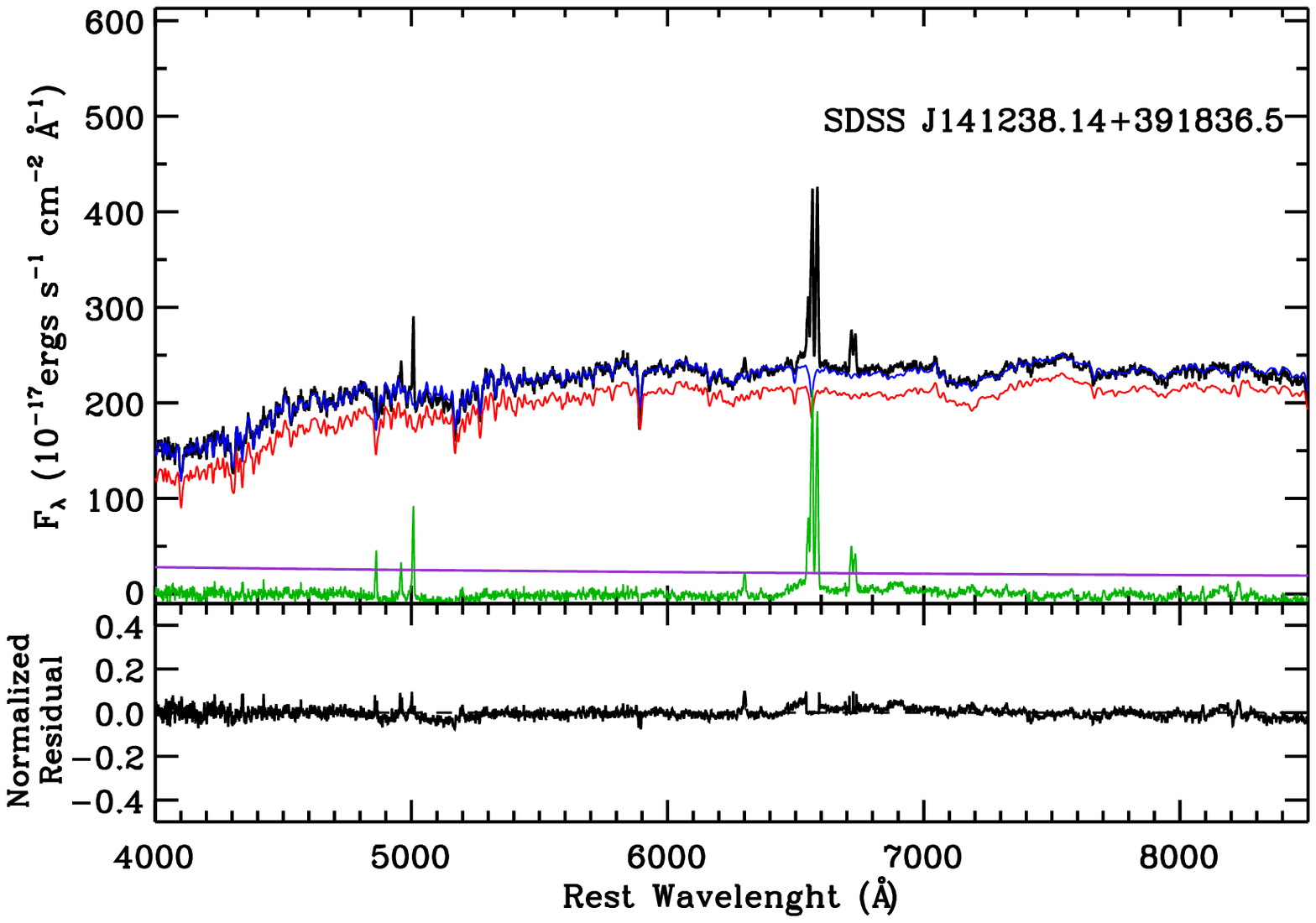}
\caption{Spectra of  J121607.08+504930.0 and J141238.14+391836.5. As in Fig.~\ref{fig:stl1}.}
\label{fig:stl3}
\end{figure*}

\begin{figure*}
\includegraphics[angle=0,width=7cm,height=7.5cm]{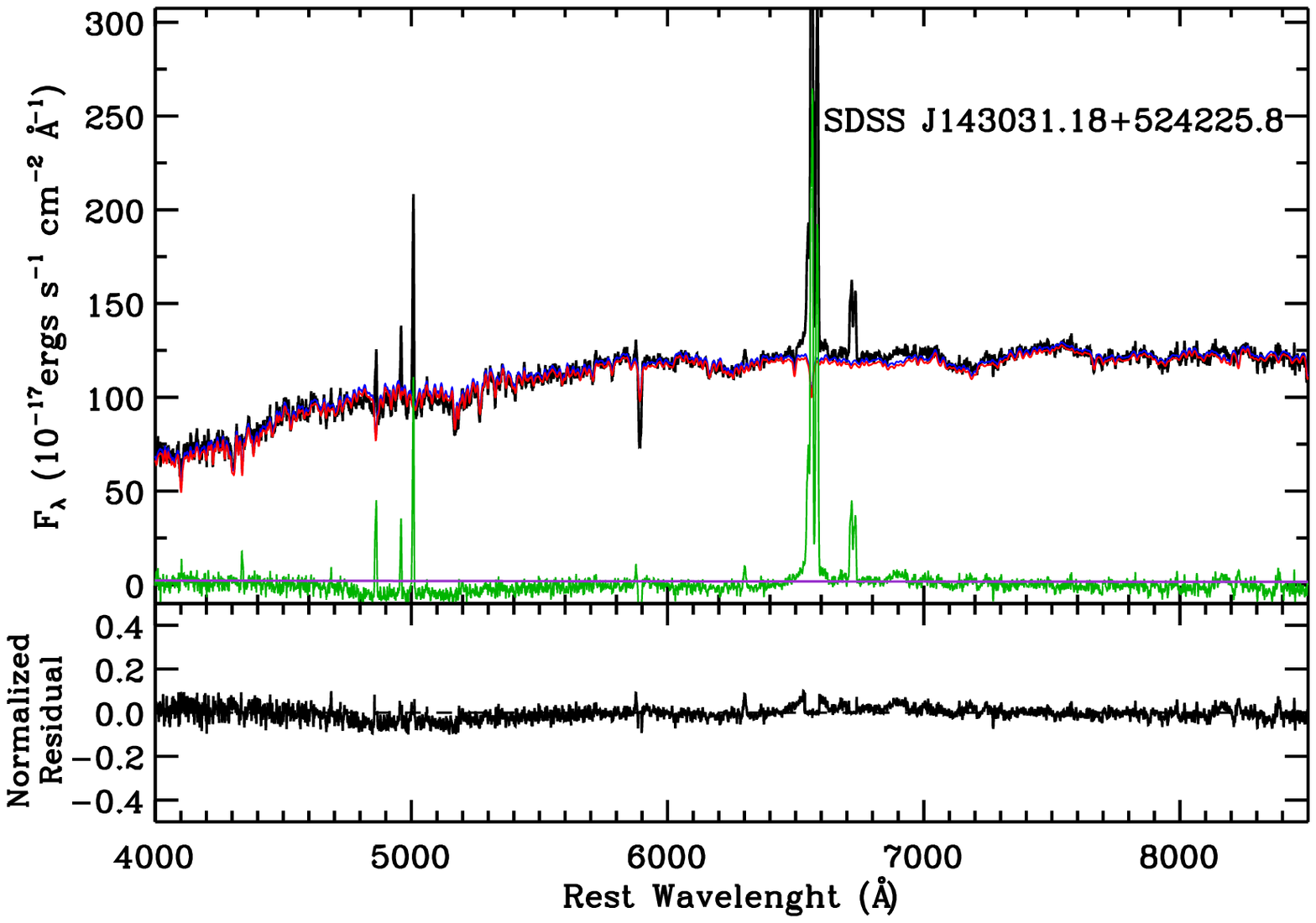}\hspace*{\columnsep}
\includegraphics[angle=0,width=7cm,height=7.5cm]{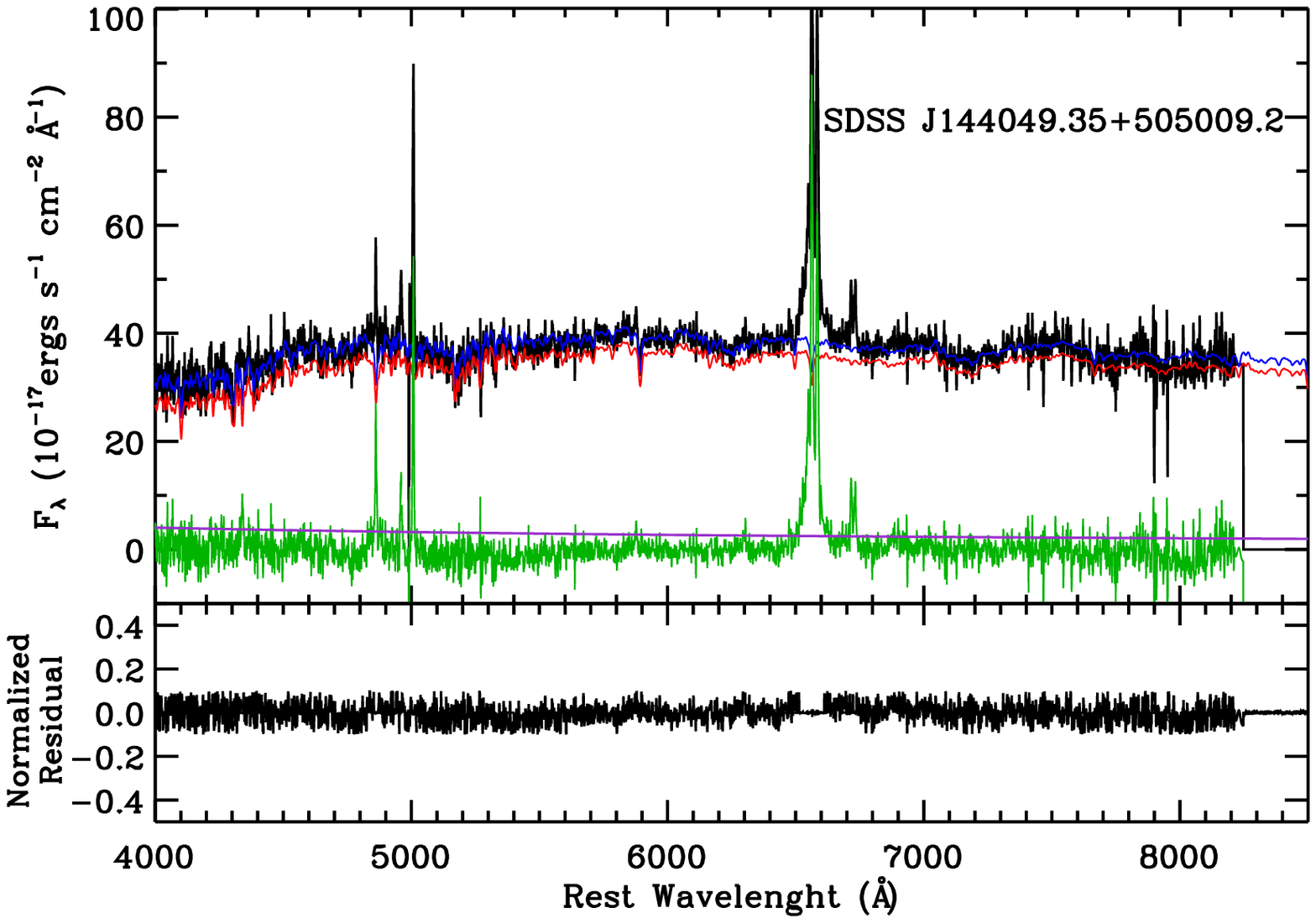}
\caption{Spectra of J143031.18+524225.8 and  J144049.35+505009.2. As in Fig.~\ref{fig:stl1}.}
\label{fig:stl5}
\end{figure*}

\begin{figure*}
\includegraphics[angle=0,width=7cm,height=7.5cm]{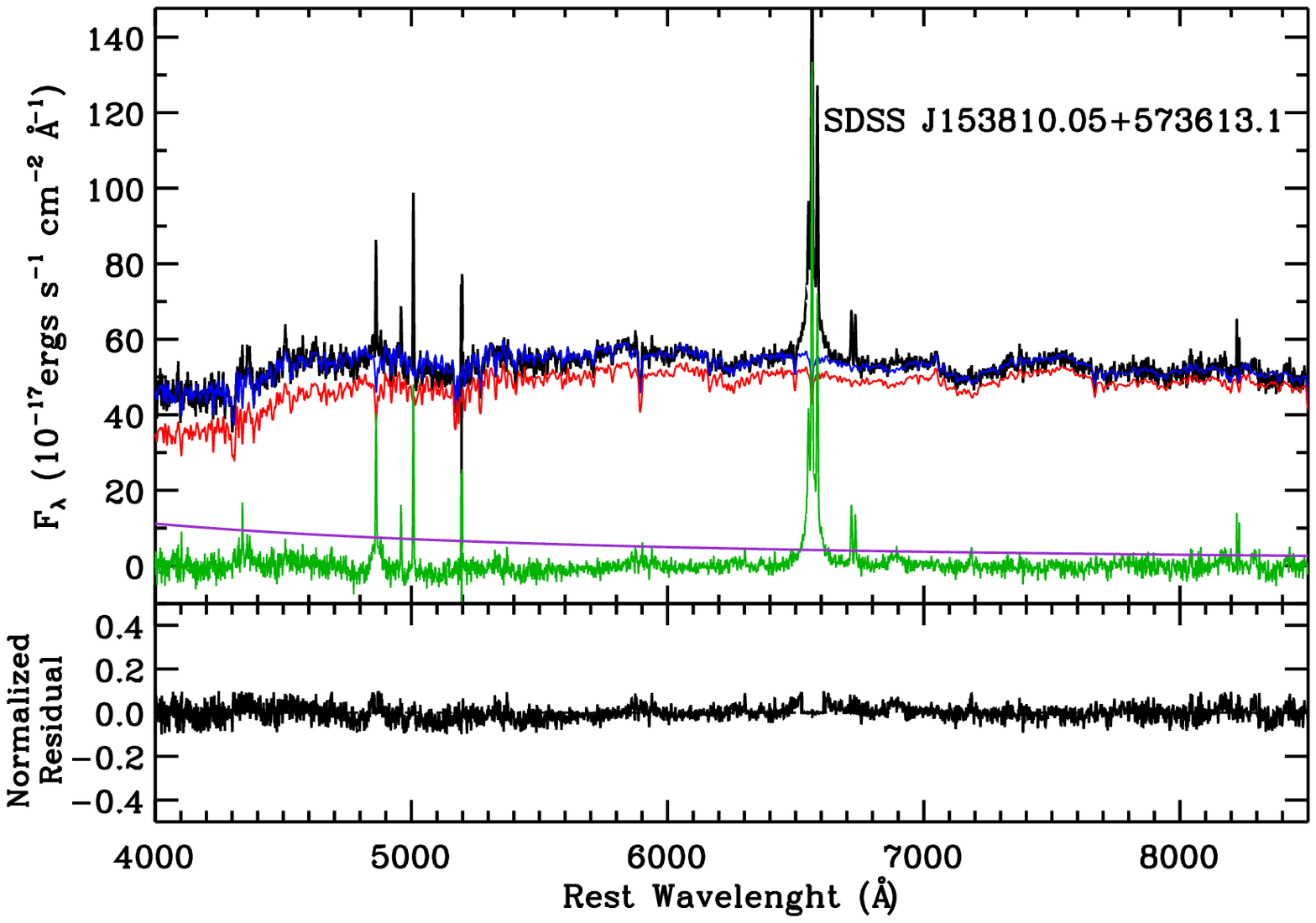}\hspace*{\columnsep}
\includegraphics[angle=0,width=7cm,height=7.5cm]{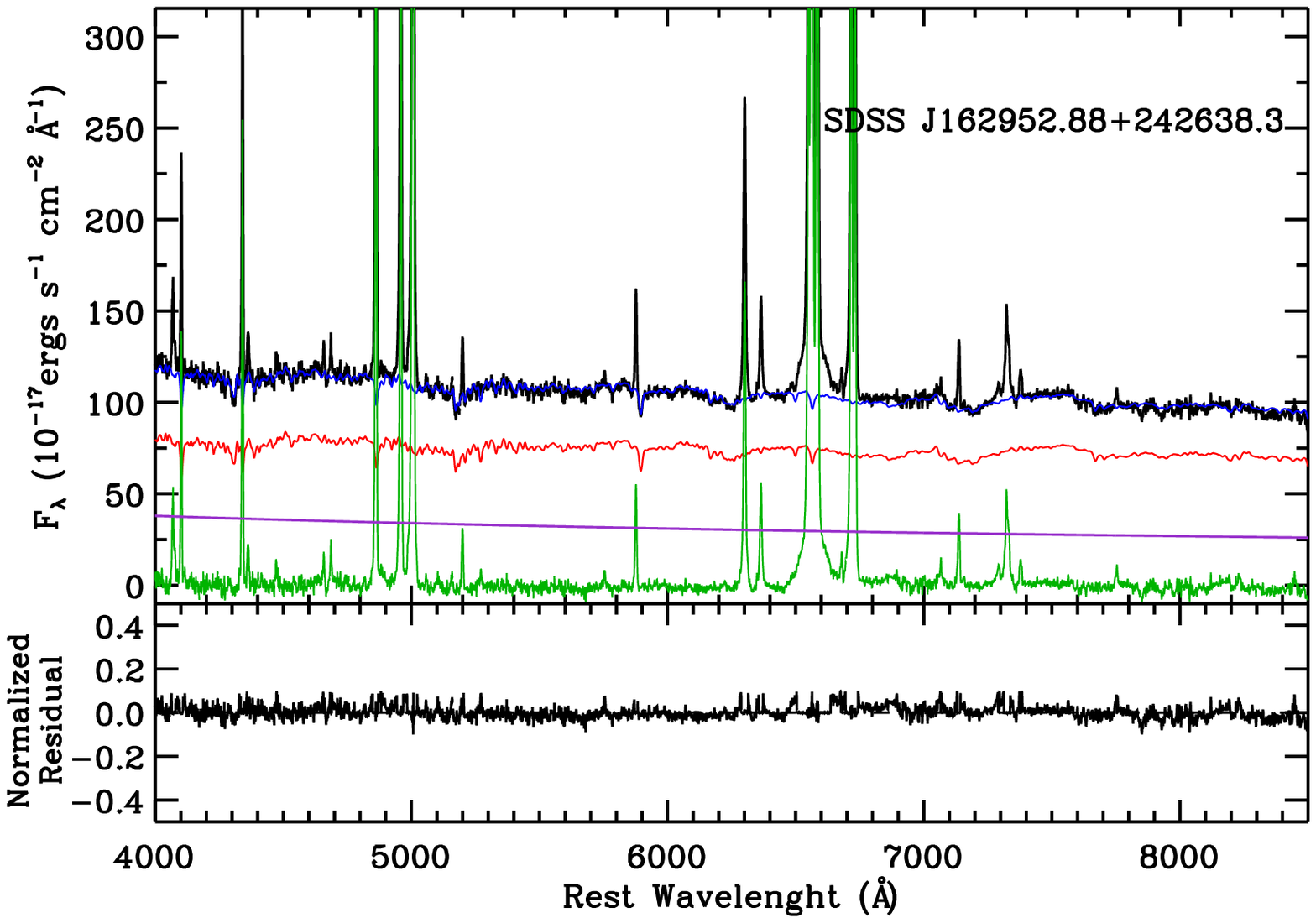}
\caption{Spectra of J153810.05+573613.1 and J162952.88+242638.3. As in Fig.~\ref{fig:stl1}.}
\label{fig:stl7}
\end{figure*}

\begin{figure*}
\includegraphics[angle=0,width=7cm,height=7.5cm]{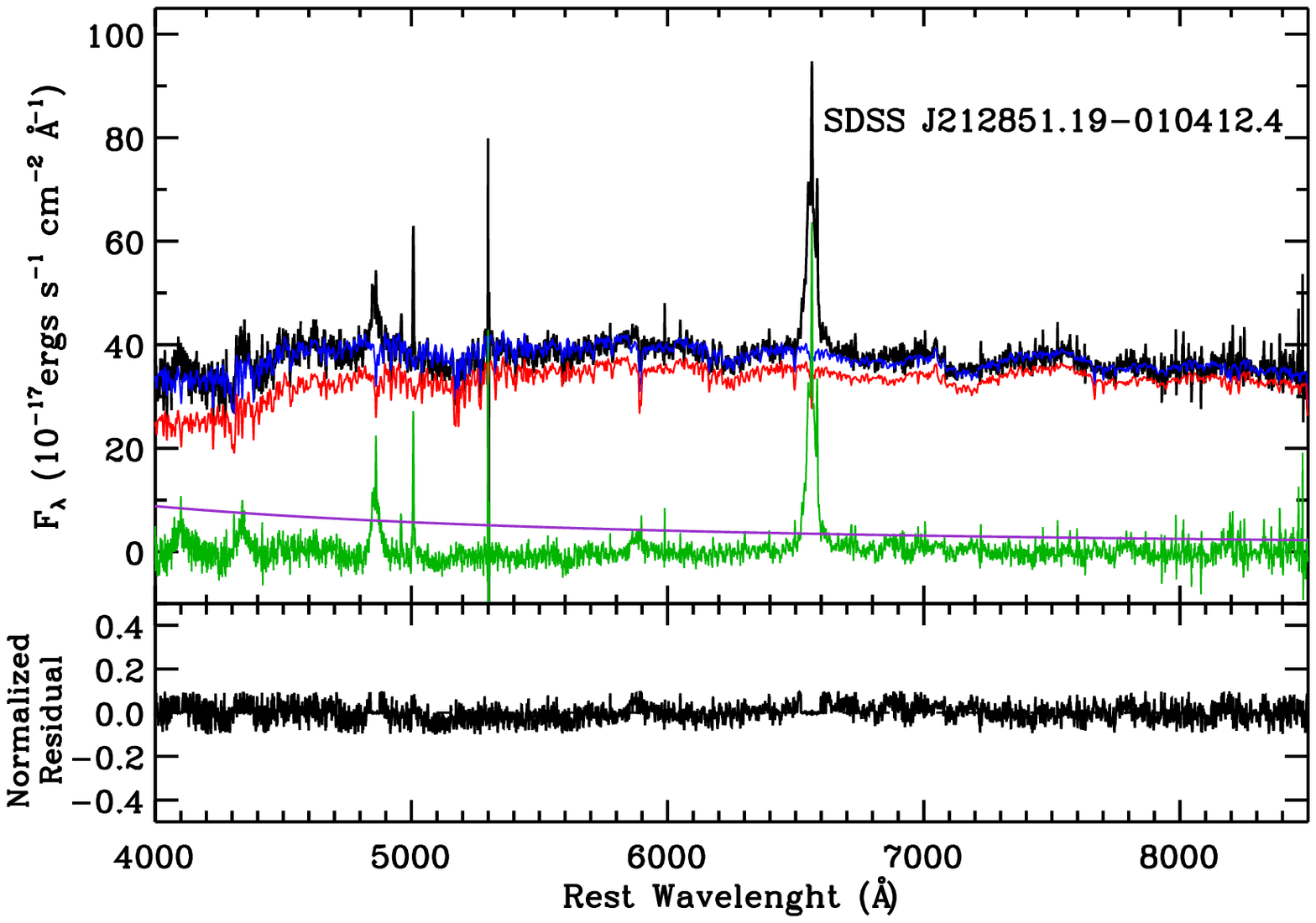}\hspace*{\columnsep}
\includegraphics[angle=0,width=7cm,height=7.5cm]{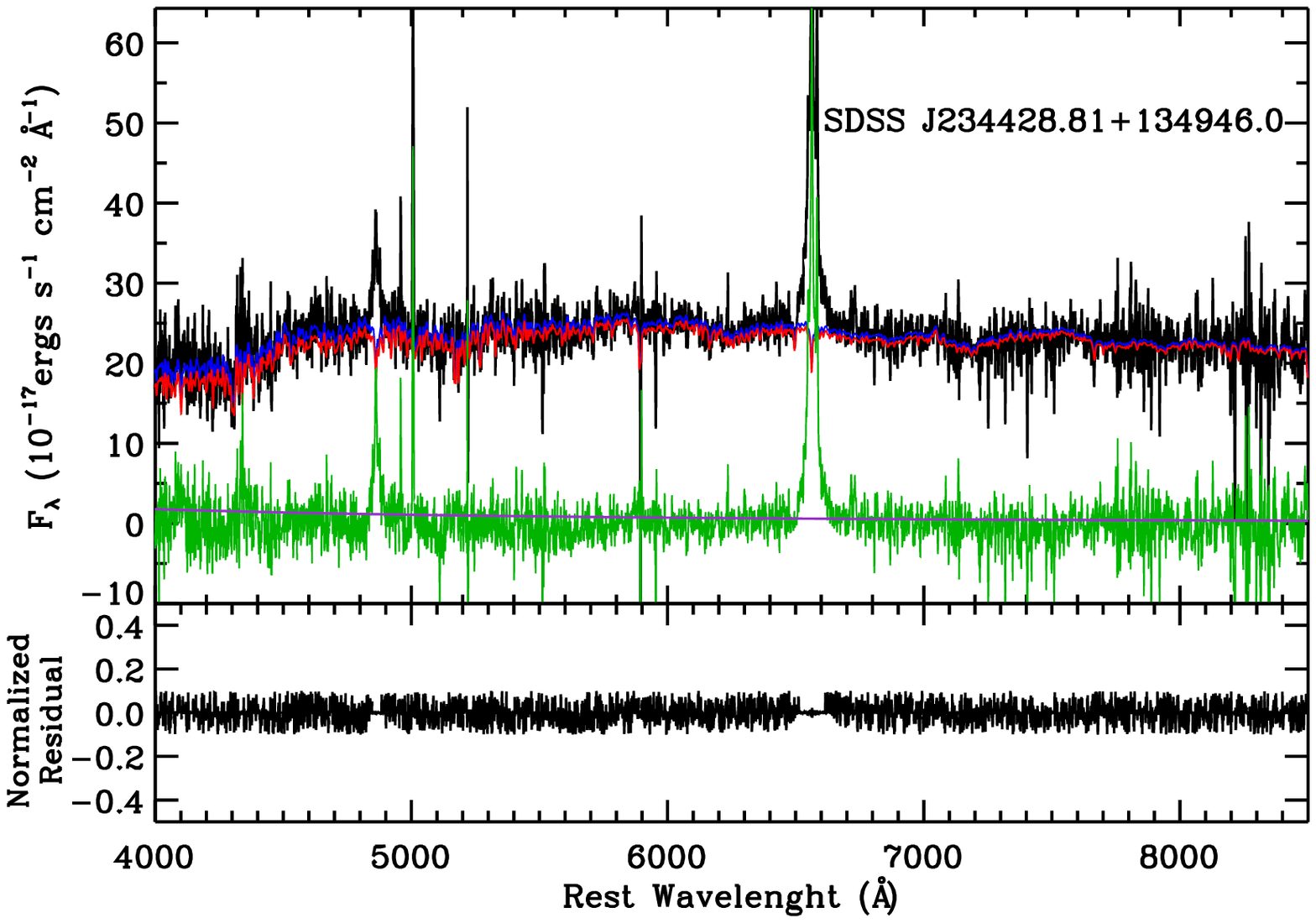}
\caption{Spectra of J212851.19-010412.4 and J234428.81+134946.0. As in Fig.~\ref{fig:stl1}.}
\label{fig:stl9}
\end{figure*}

{\bf
\begin{figure*}
{\includegraphics[angle=90,width=\textwidth]{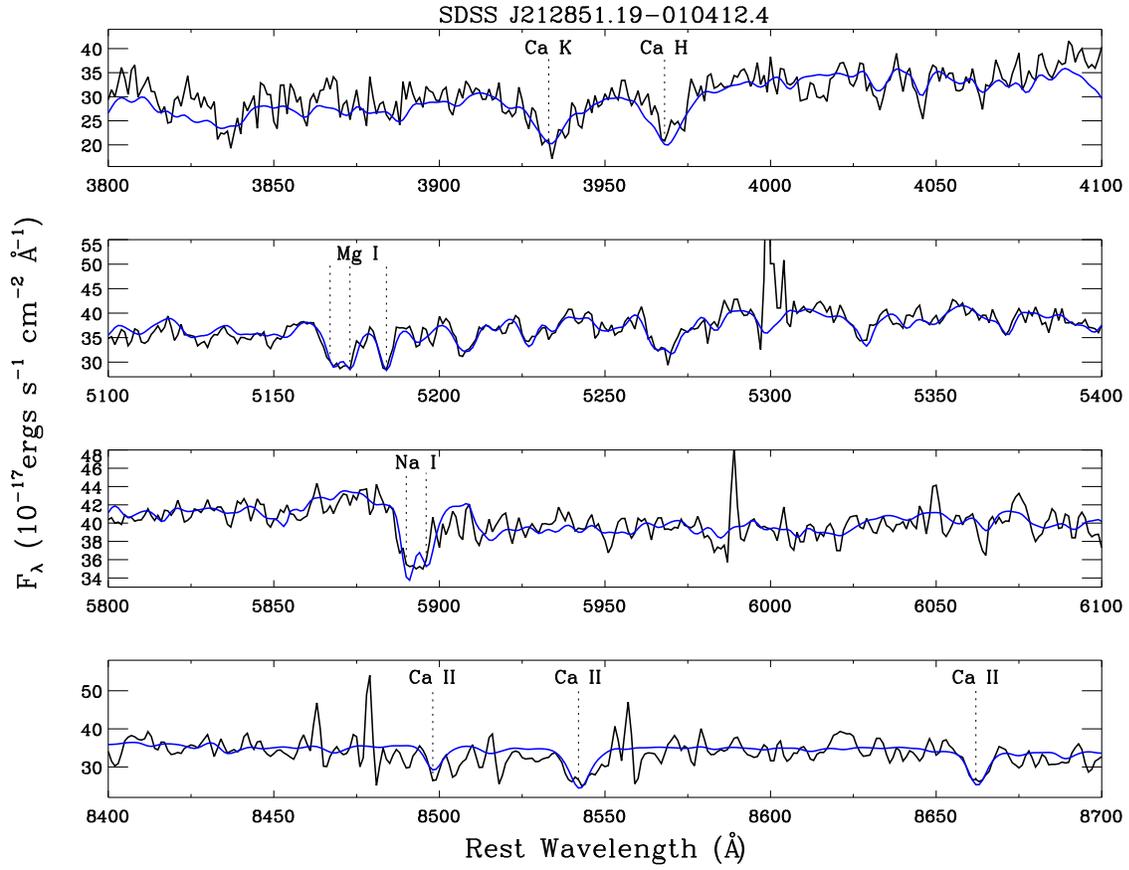}}
\caption{A blow-up of our best model obtained from Monte Carlo simulations for the the spectrum of 
J212851.19-010412.4. This fit is shown also in blue in the left panel of Fig.~\ref{fig:stl9}. 
The lines used to estimate the velocity dispersion are marked with vertical dotted lines.}
\label{fig:best}
\end{figure*}
}

\begin{figure*}
\includegraphics[width=6cm, angle=90]{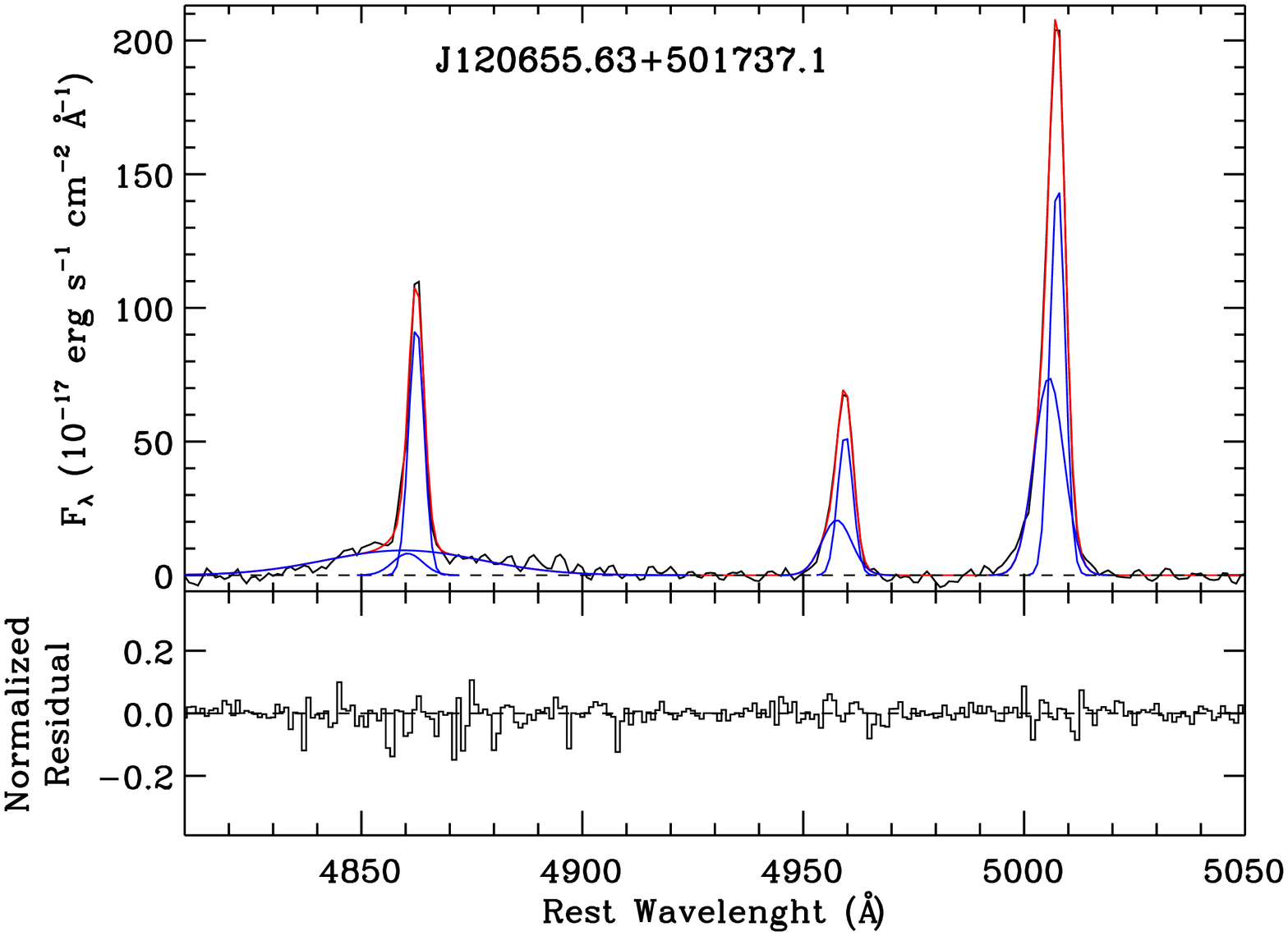}\hspace*{\columnsep}
\includegraphics[width=6cm, angle=90]{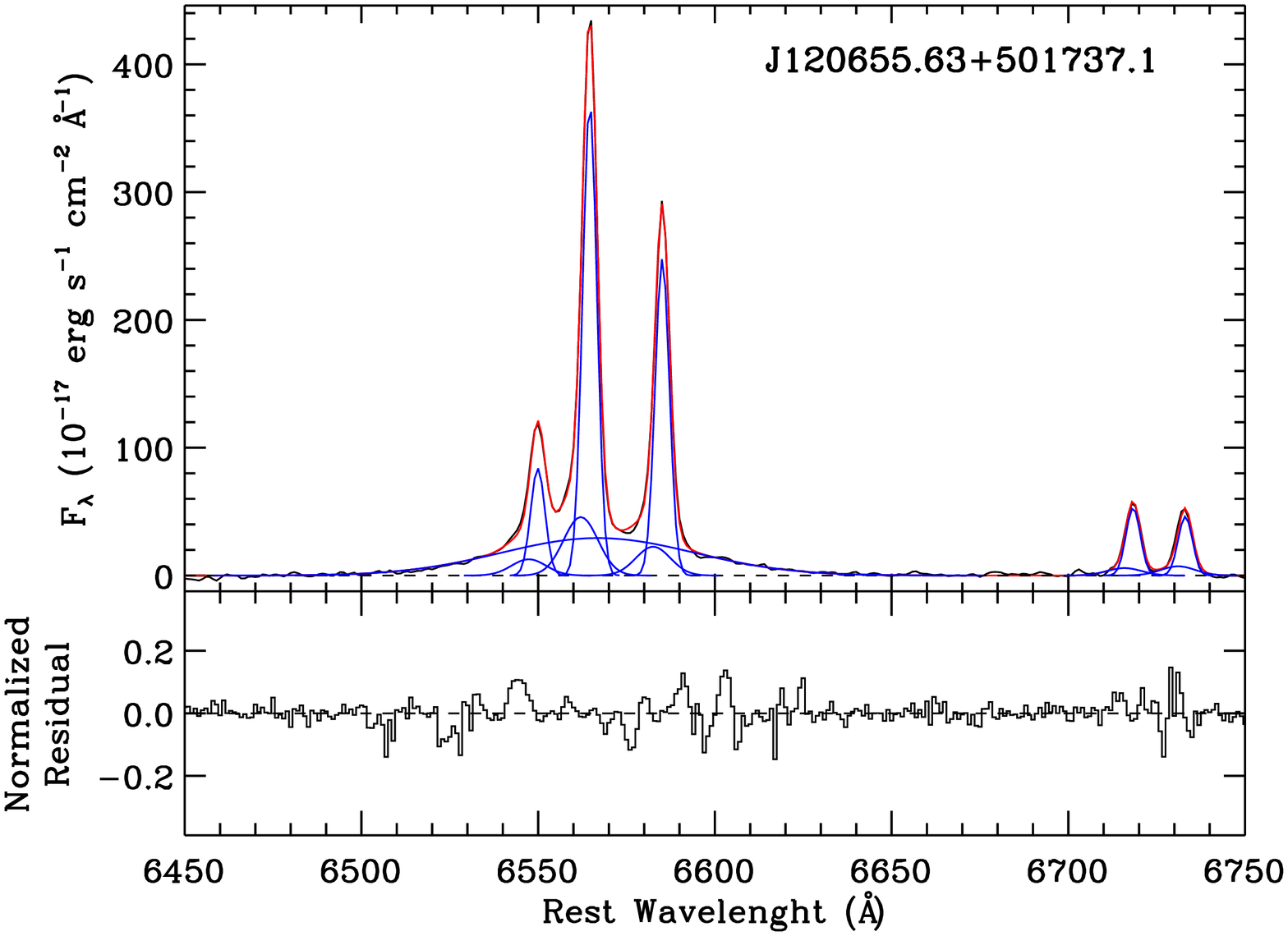}
\caption{Spectral decomposition for SDSS J120655.63+501737.1. Left  
and right panels show our best fit obtained for the blue and red part 
of the spectrum, respectively. 
In both panels black lines correspond to the pure AGN spectrum obtained 
with STARLIGHT, blue lines show the Gaussian components used for the 
fit and red lines the best fit obtained. At the bottom of each panel 
the corresponding normalized residuals are shown. The  {\textsc{[O\,III]}}$\lambda$4959,5007 
lines and the narrow component of H$\beta$ were best fitted using 2 
Gaussian components. For the broad component of H$\beta$ only one Gaussian 
was needed for the fit. The red part of the spectrum was modeled in a 
similar way, i.e., two Gaussian components for the narrow lines and one 
Gaussian for the broad component of H$\alpha$.} 
\label{fig:dblpk1}
\end{figure*}

\begin{figure*}
\includegraphics[width=6cm,angle=90]{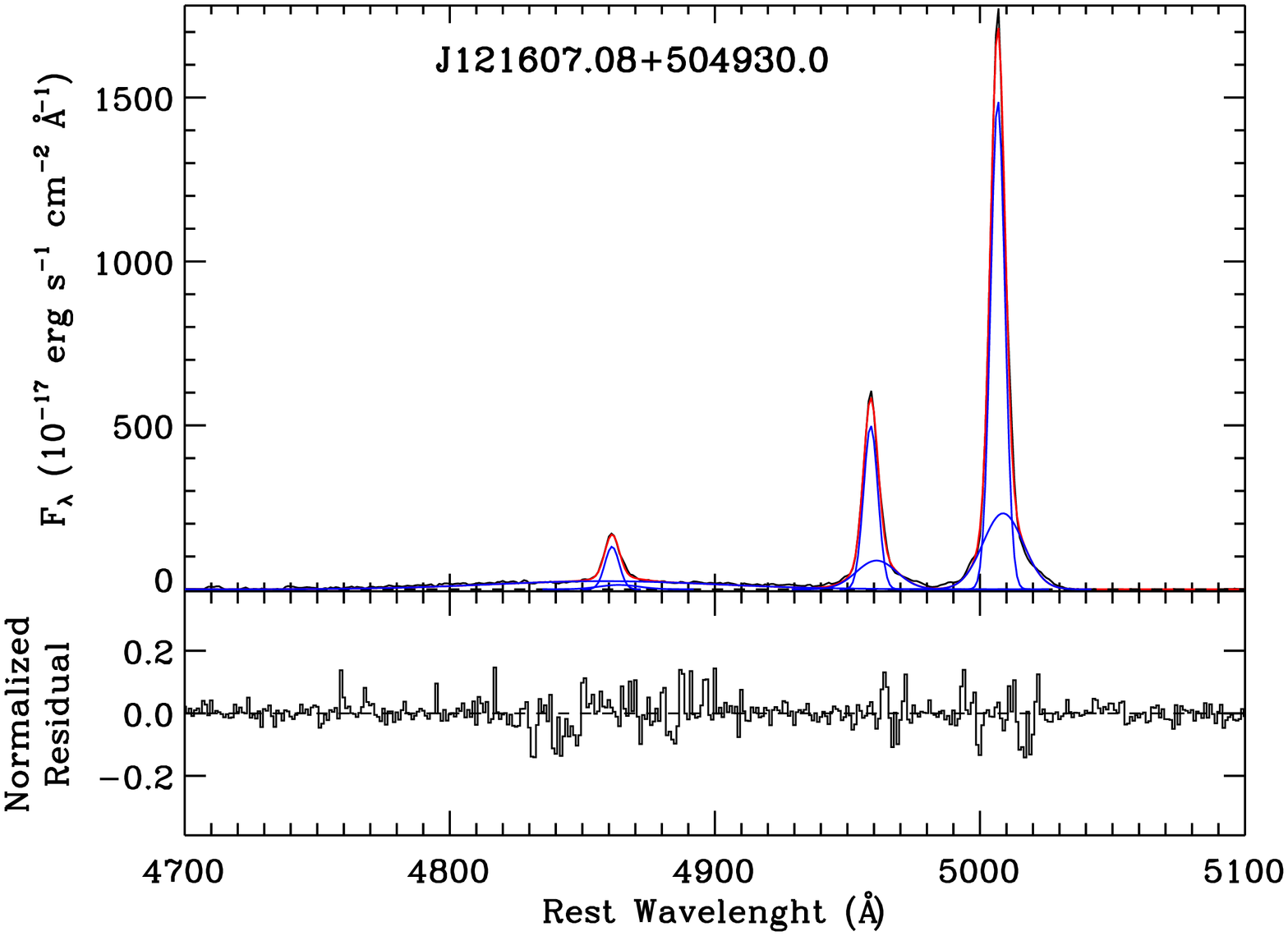}\hspace*{\columnsep}
\includegraphics[width=6cm,angle=90]{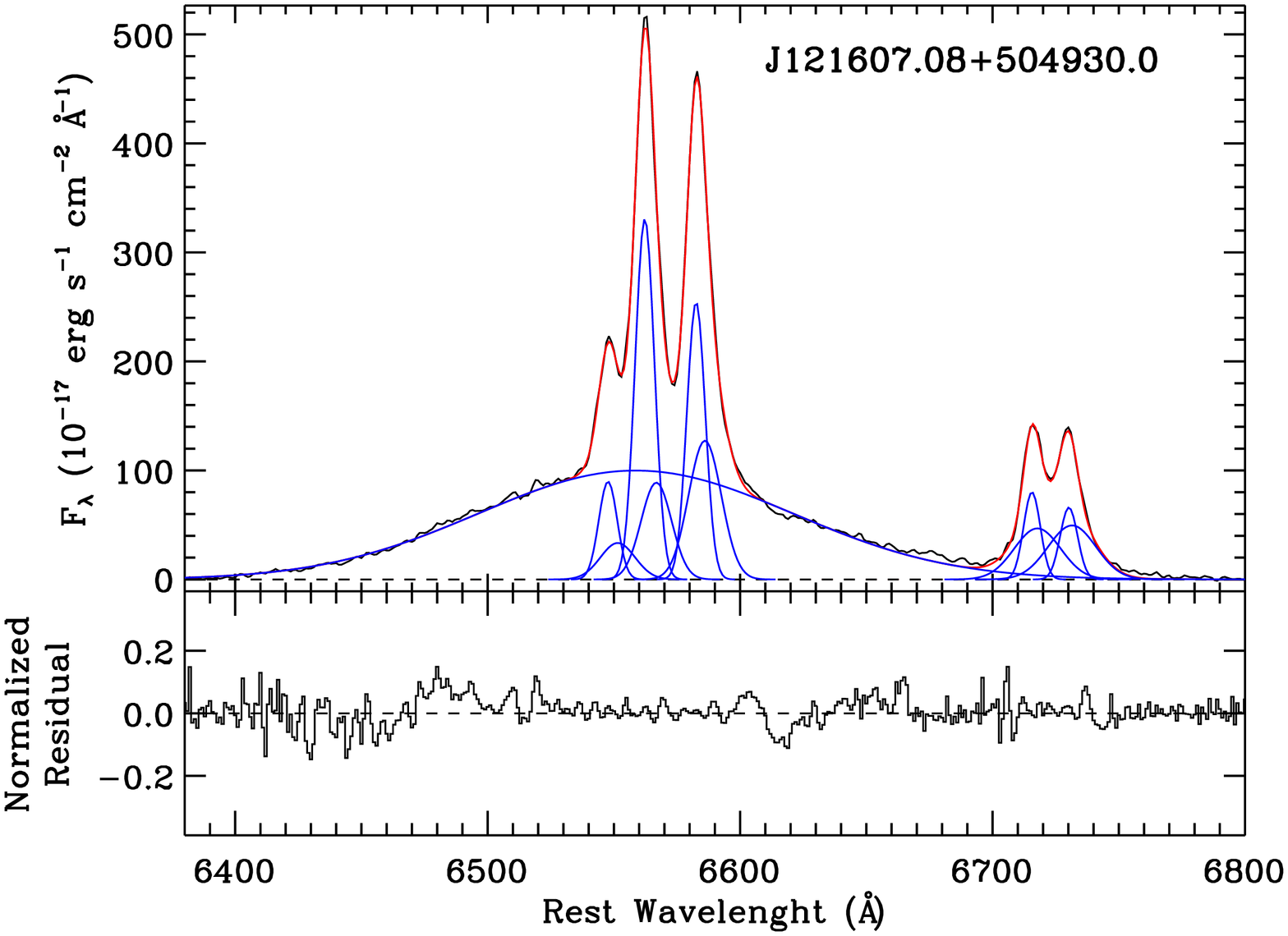}
\caption{Spectral decomposition for SDSS J121607.08+504930.0. We have modeled the lines 
in a similar way as described in Fig.~\ref{fig:dblpk1}.}
\label{fig:dblpk3}
\end{figure*}

\begin{figure*}
\includegraphics[width=6cm, angle=90]{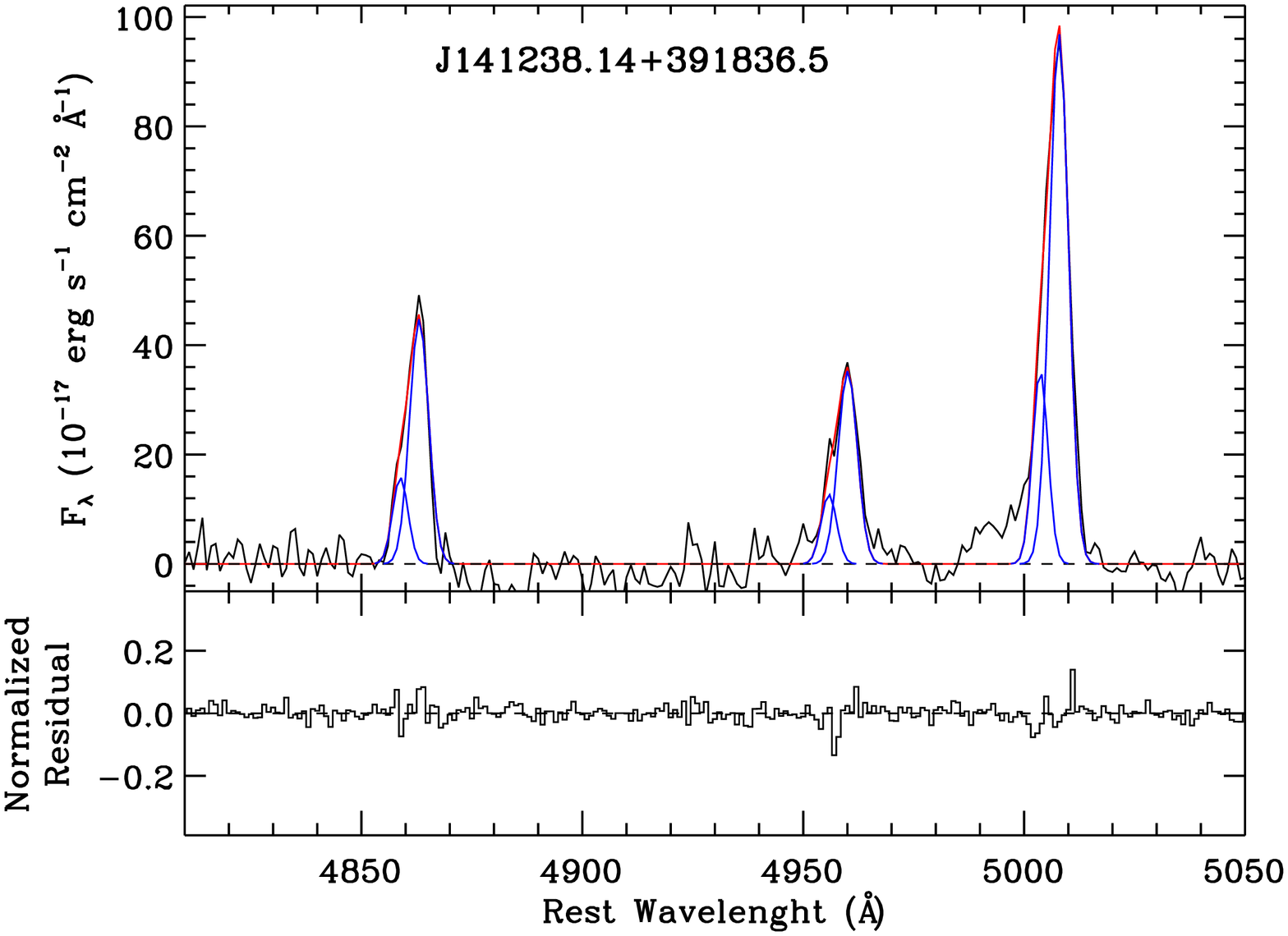}\hspace*{\columnsep}
\includegraphics[width=6cm, angle=90]{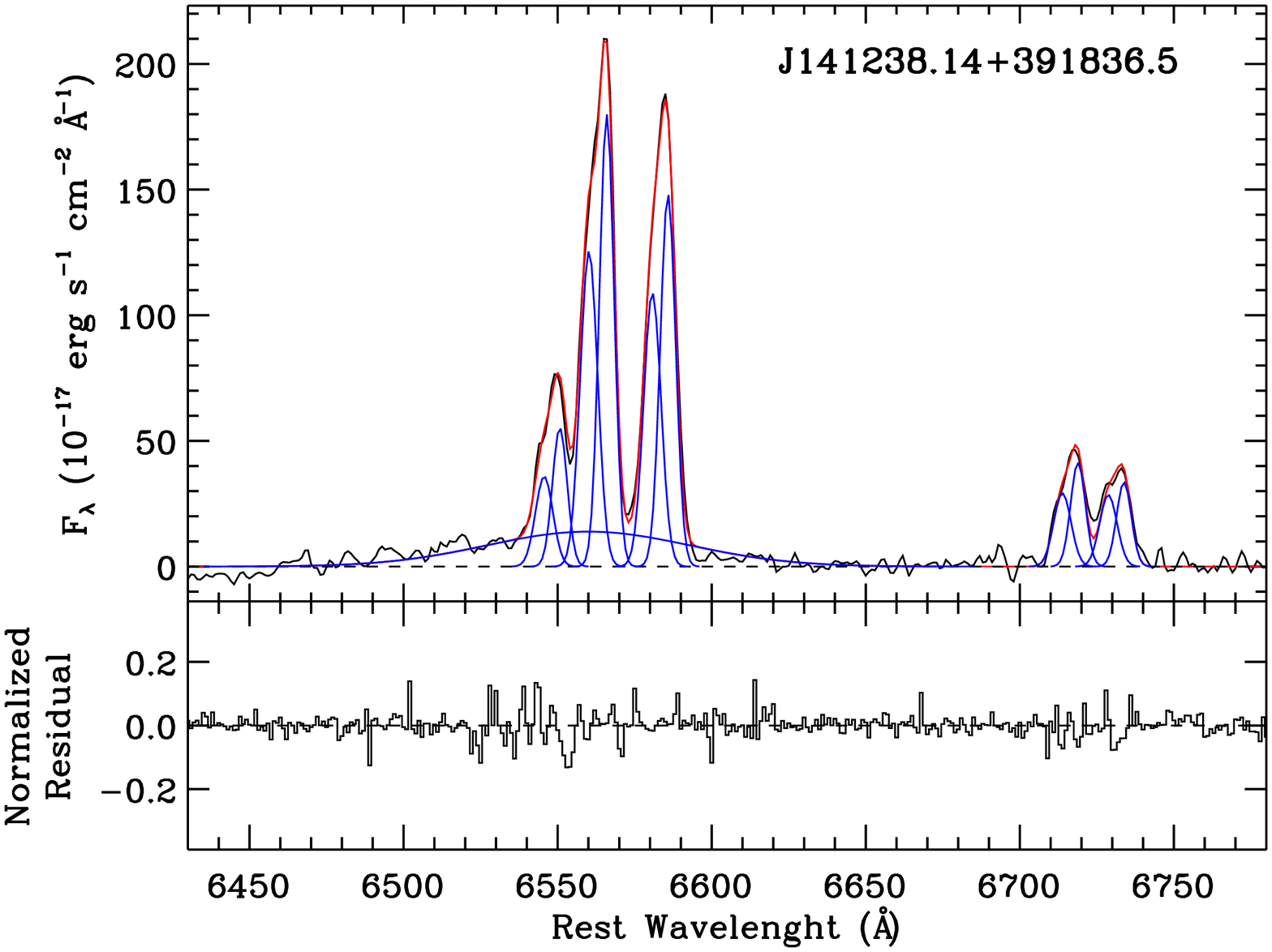}
\caption{Spectral decomposition for SDSS J141238.14+391836.5. This object 
has also been modeled using two Gaussian components for the narrow lines. But, 
when fitting the broad components we found  no evidence 
for a broad component in H$\beta$. For the red part of the spectrum a broad 
Gaussian component was needed for properly fitting 
the H$\alpha$ line, and the two Gaussian components for the narrow lines.} 
\label{fig:dblpk4}
\end{figure*}

\begin{figure*}
\includegraphics[width=6cm,angle=90]{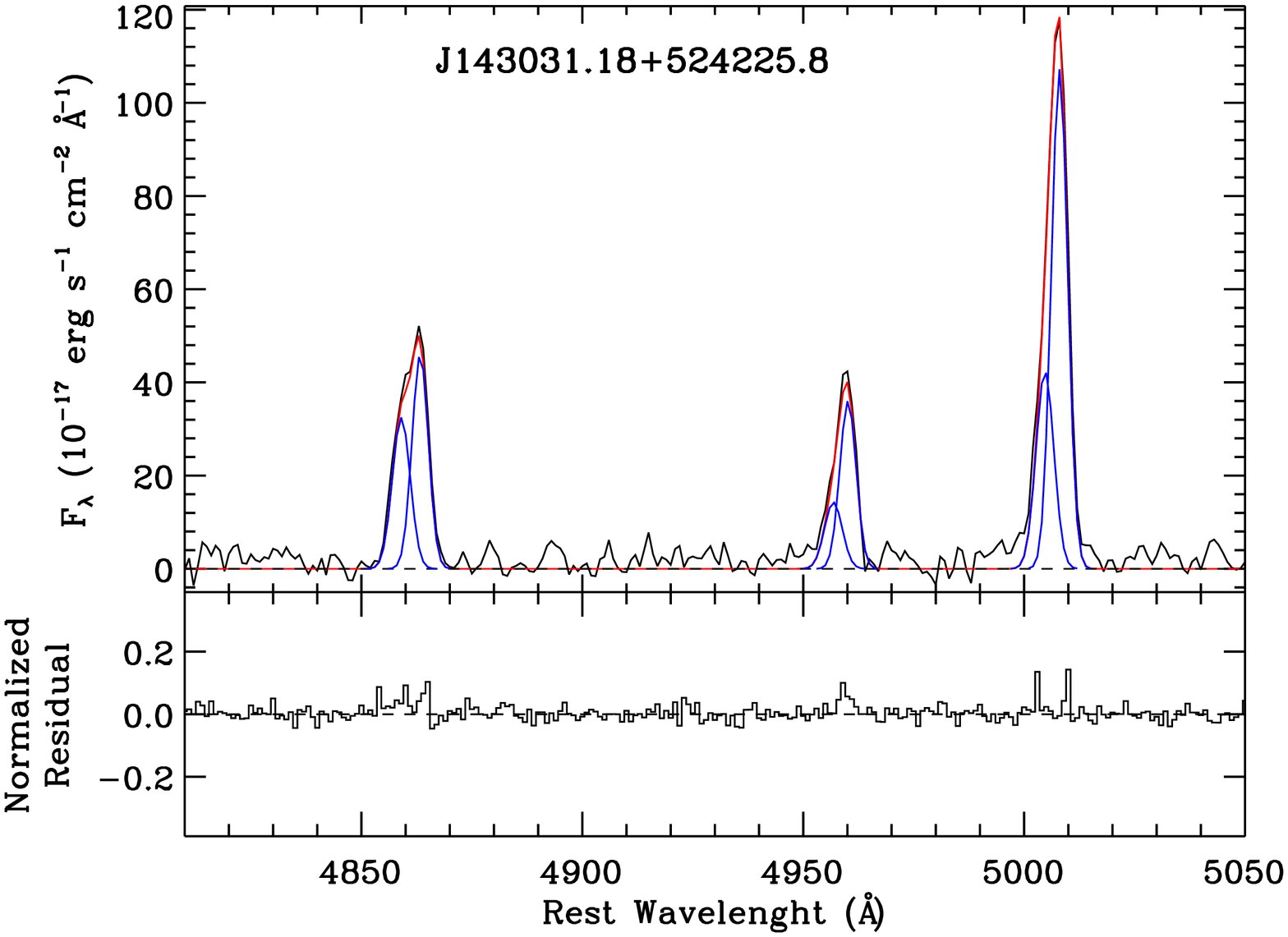}\hspace*{\columnsep}
\includegraphics[width=6cm, angle=90]{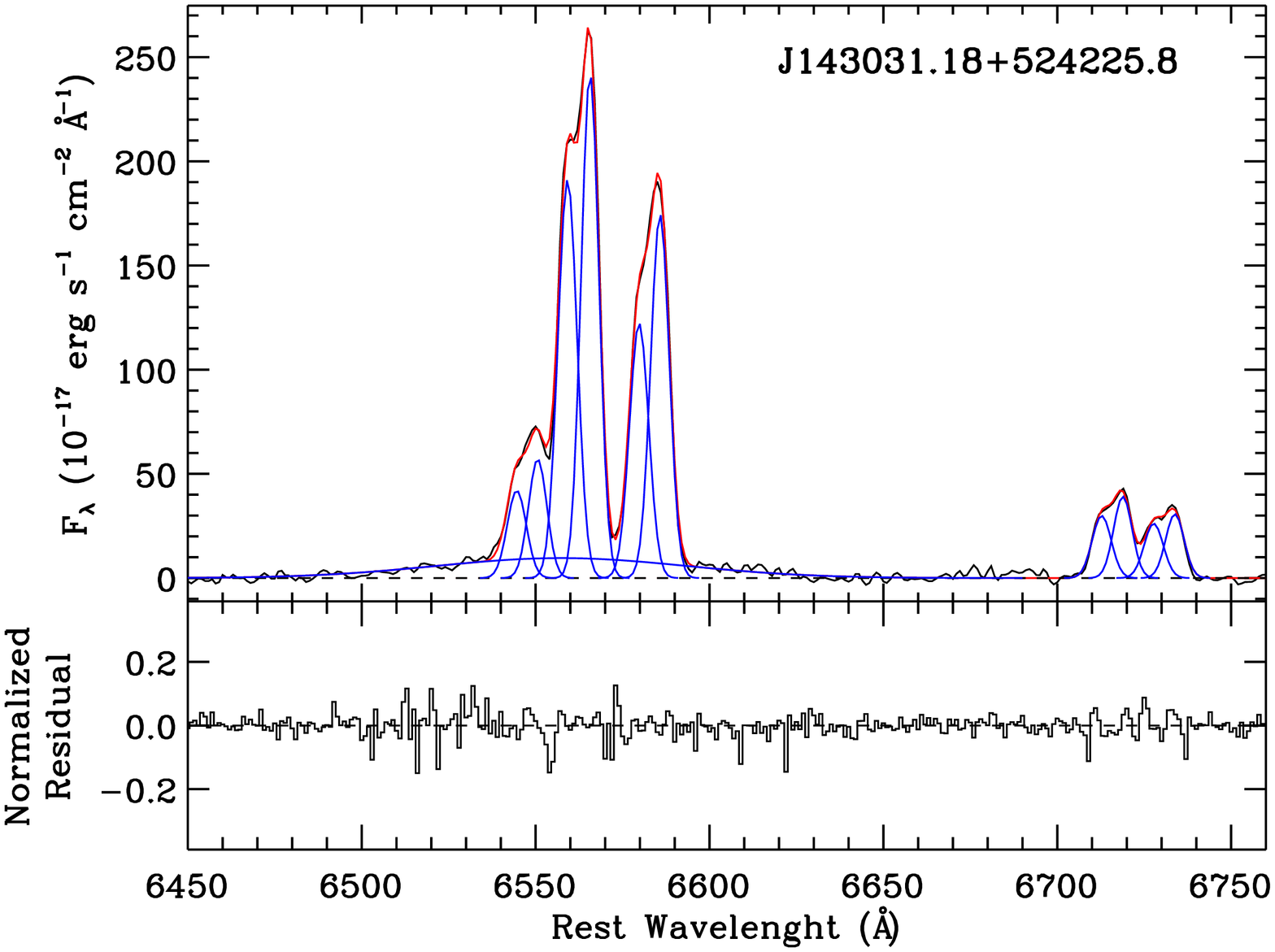}
\caption{Spectral decomposition for SDSS J143031.18+524225.8. This object 
has been modeled similar to the one shown in  Fig.~\ref{fig:dblpk4}.
} 
\label{fig:dblpk5}
\end{figure*}

\begin{figure*}
\includegraphics[width=6cm, angle=90]{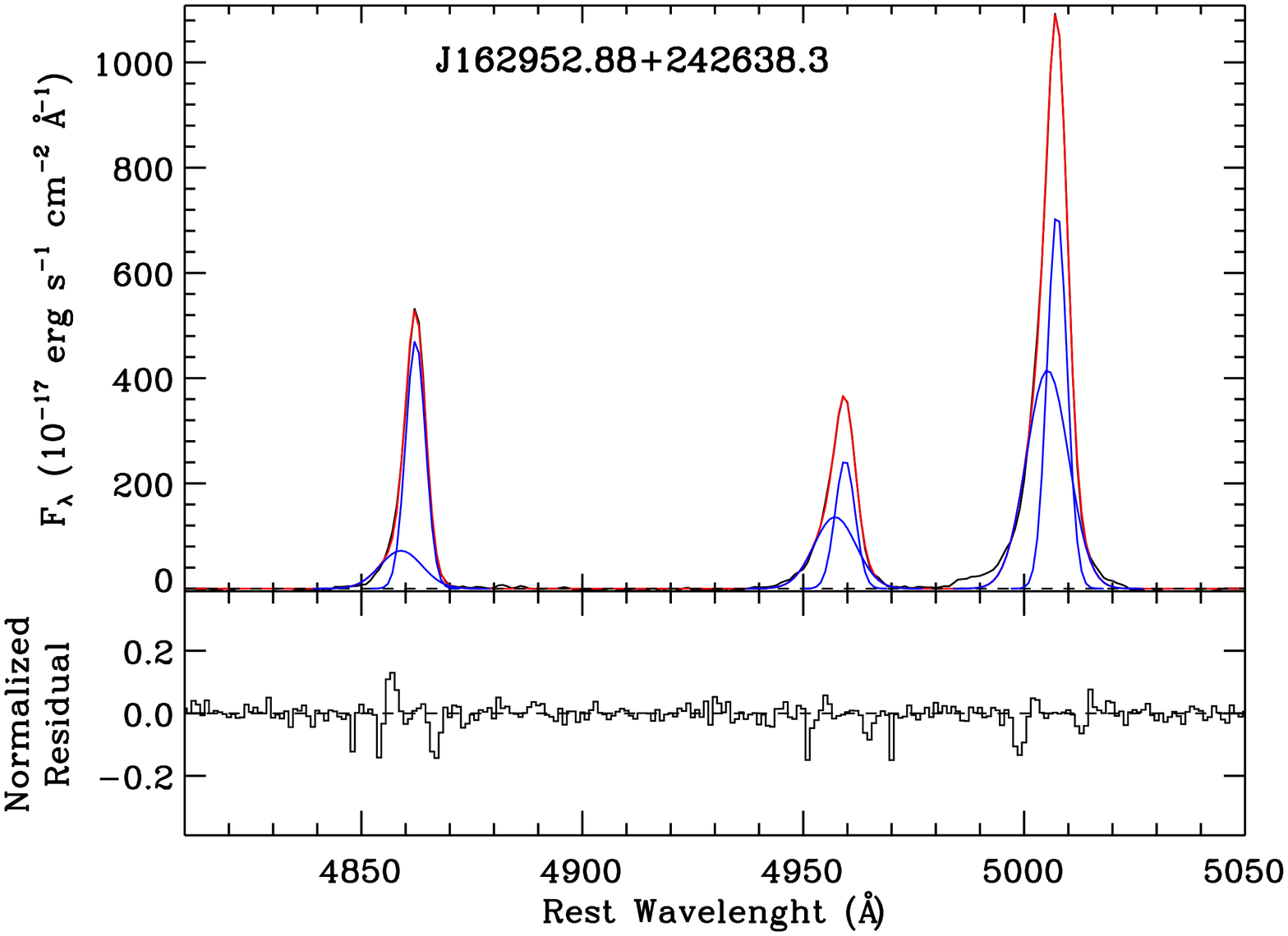}\hspace*{\columnsep}
\includegraphics[width=6cm, angle=90]{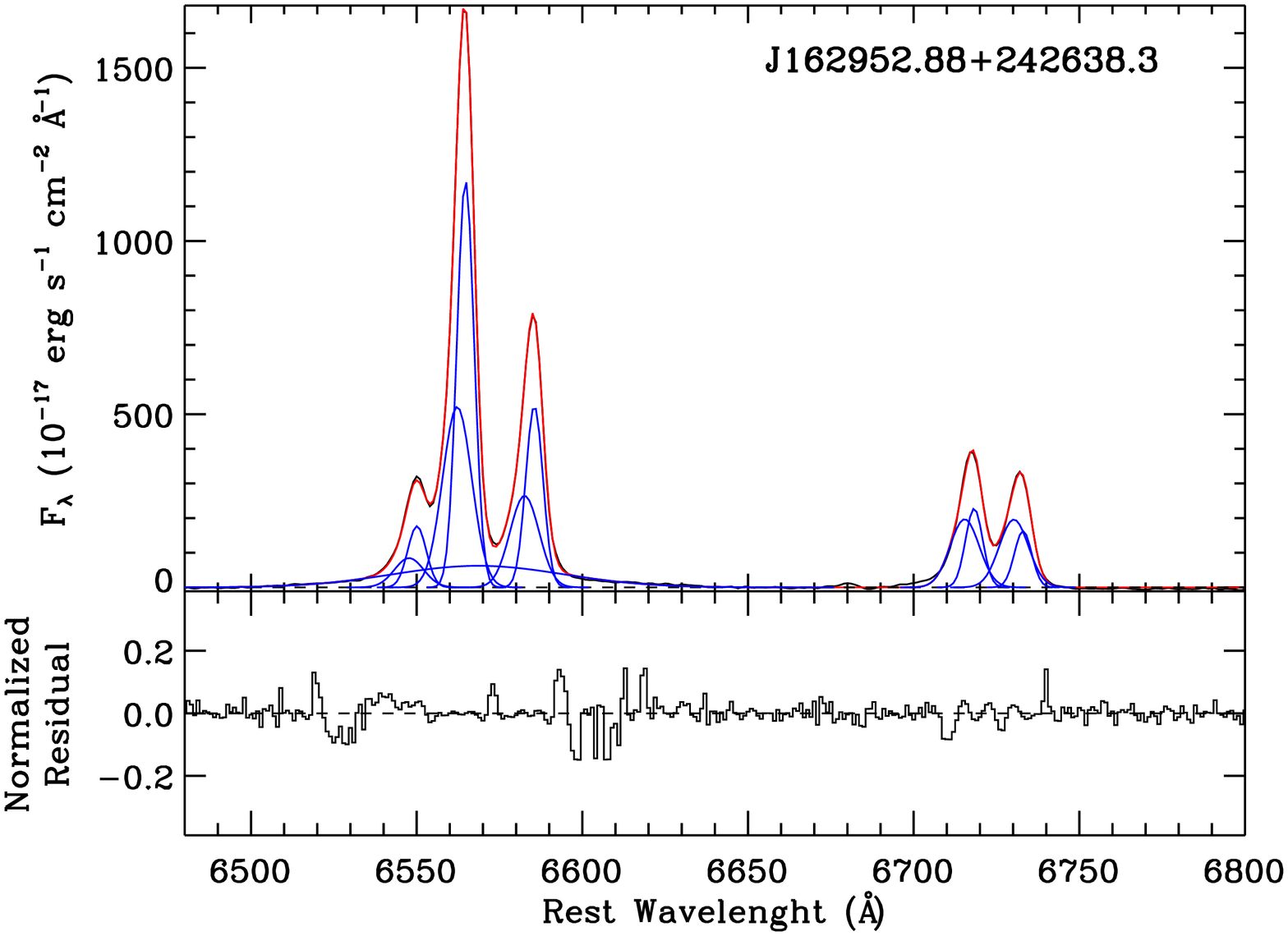}
\caption{Spectral decomposition for SDSS J162952.88+242638.3. The lines 
were fitted using three Gaussian components for {\textsc{[O\,III]}}$\lambda$4959,5007 
since it shows a wing component in addition to the two narrow components.
The red part of the spectrum was fitted using two Gaussian components for the
narrow lines. There is no broad component in H$\beta$, but a broad component
was fitted for H$\alpha$.
} 
\label{fig:dblpk8}
\end{figure*}

\begin{figure*}
\includegraphics[width=6cm, angle=90]{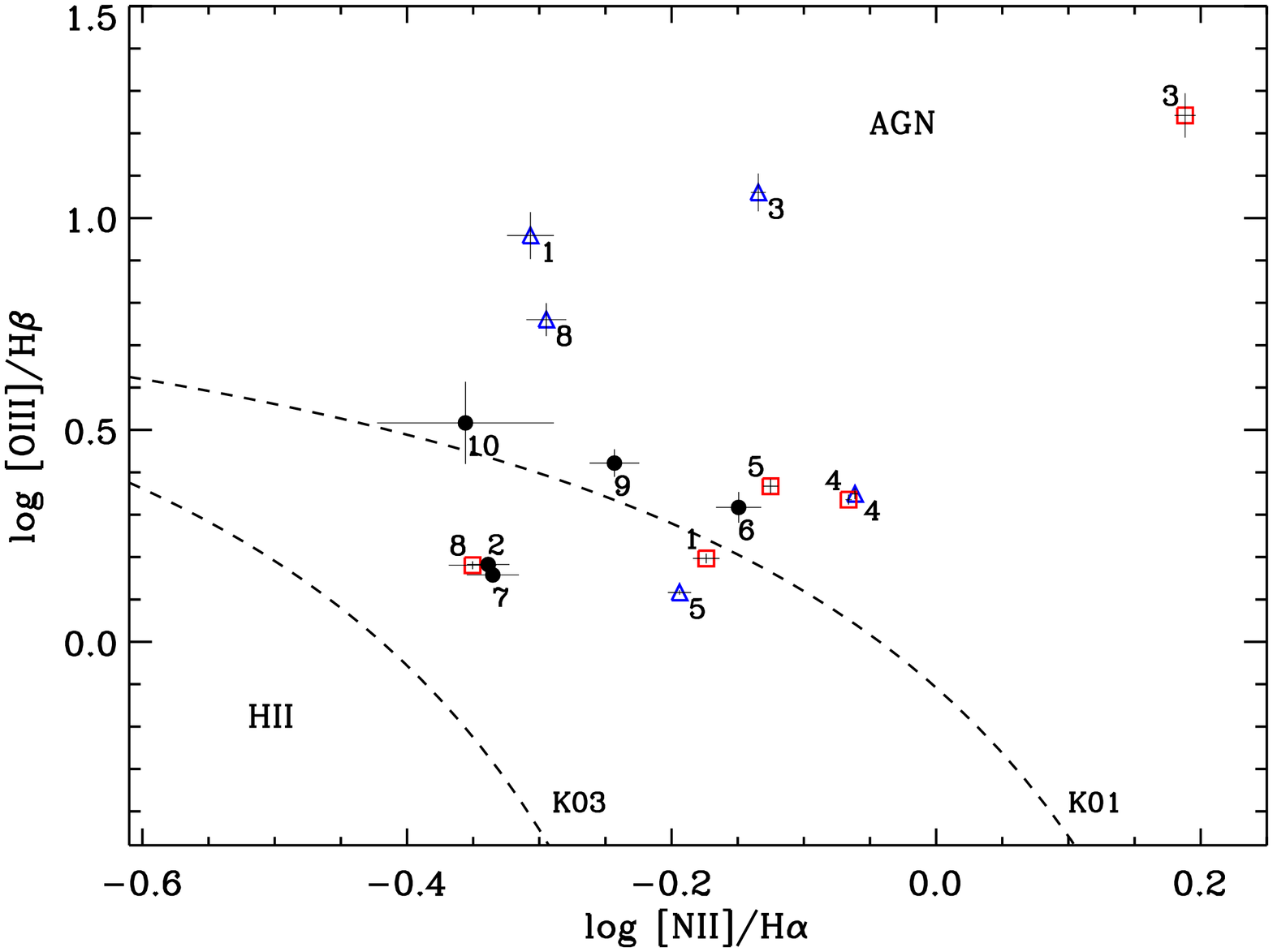}\hspace*{\columnsep}
\includegraphics[width=6cm, angle=90]{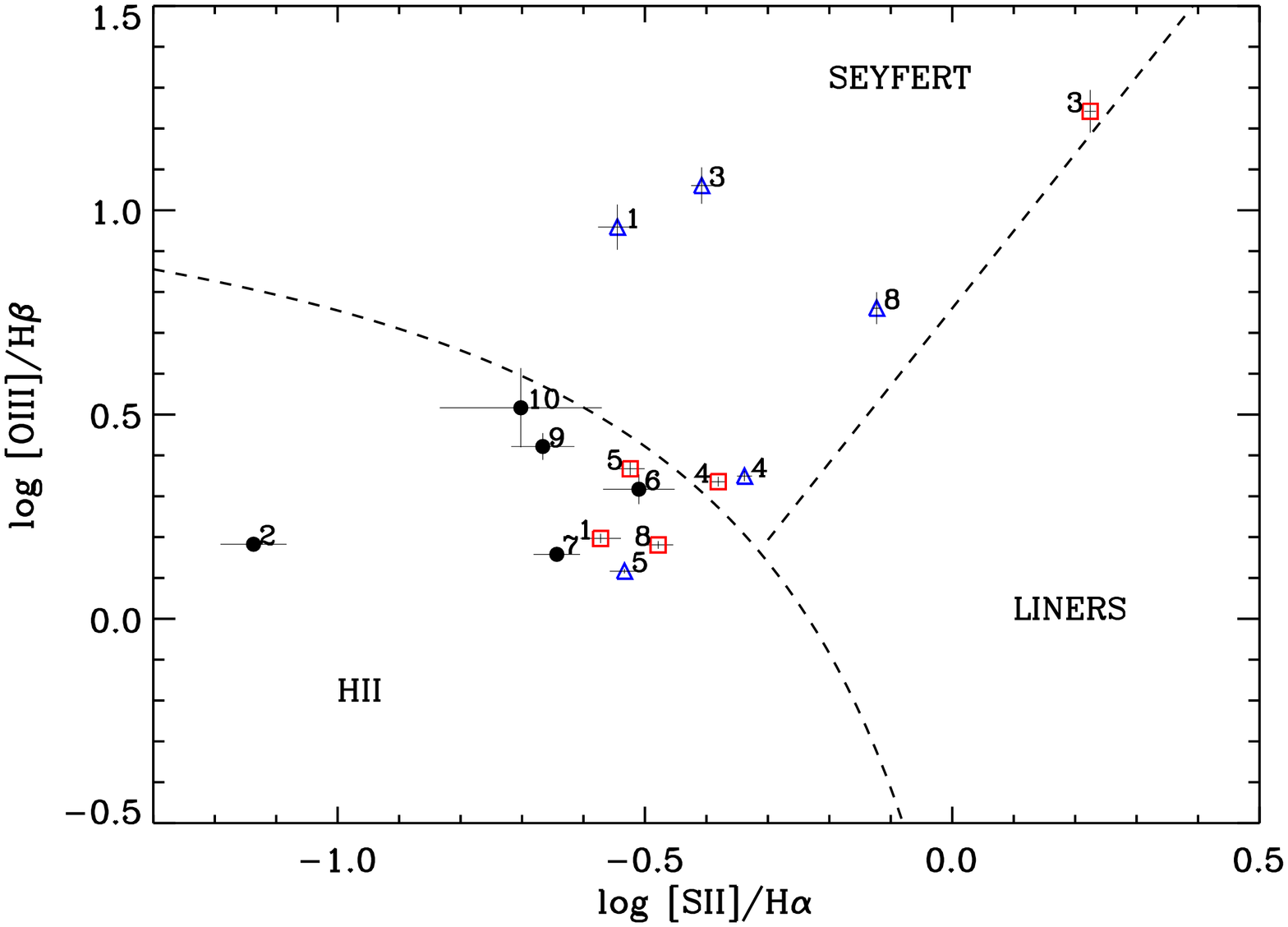}
\caption{Location of the objects of the sample in two BPT diagnostic diagrams. Left side, the dotted lines come from: \citep[][K01]{2001ApJ...556..121K} 
and \citep[][K03]{2003MNRAS.346.1055K}. Some objects are composite AGN, i.e., they lie between K01 and K03 lines, meaning that they 
have a Starburst galaxy (SB) plus an AGN. Filled black dots mark the locus of single-peaked objects.  Objects with narrow double-peaked 
emission lines are shown twice since we used the ratios obtained from their red (open squares) and blue (open triangles) components. 
Right side, BPT diagram showing that all objects are located above the LINER region accordingly with \citet{2006MNRAS.372..961K}.
The spectral classification is presented in Table~\ref{tab:BHm}.}
\label{fig:diagnostic}
\end{figure*}

\begin{figure*}
{\includegraphics[angle=90,width=\textwidth]{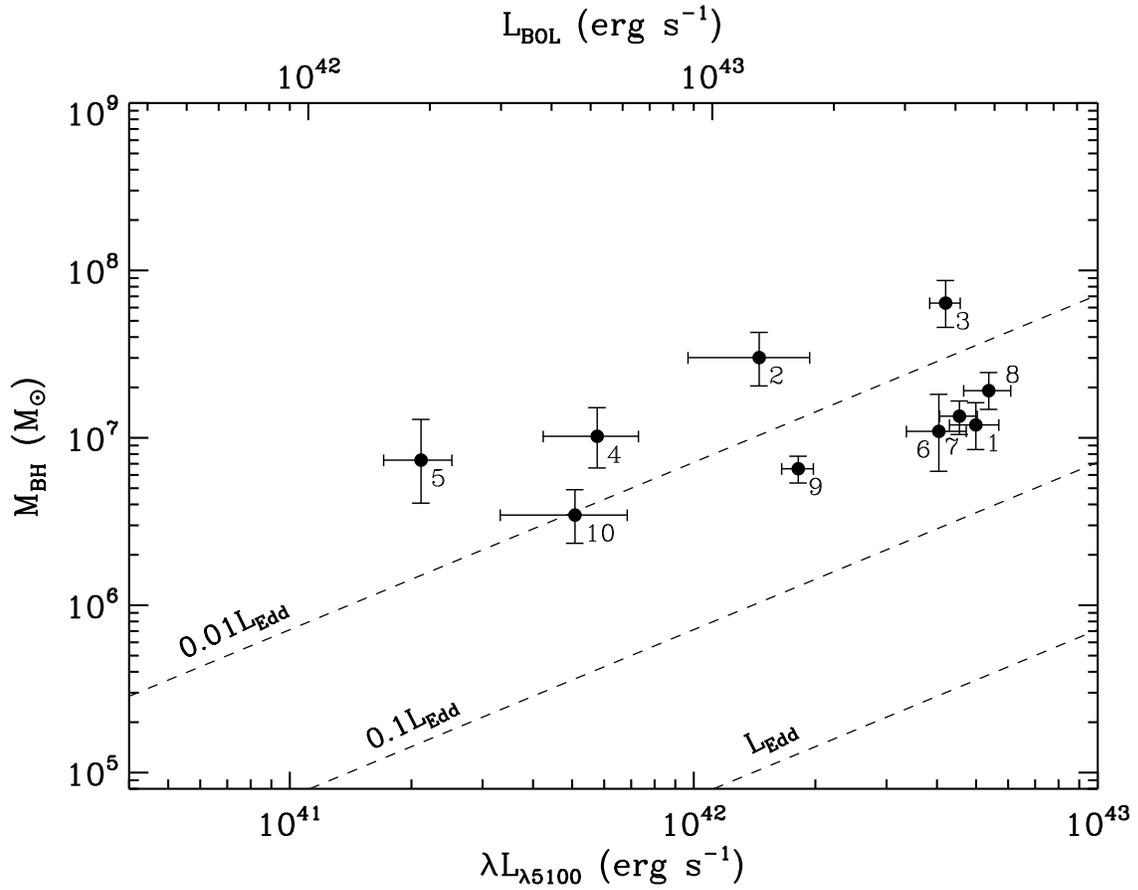}}
\caption{Diagram $M_{BH}$ vs. luminosity. The upper x-axis shows the bolometric luminosity
assuming $L_{bol}\sim$9$\lambda$$L_{\lambda5100}$.}
\label{fig:mass-lum}
\end{figure*}

\begin{figure*}
\begin{center}
\begin{tabular}{lr}
 \includegraphics[height=6.4cm,width=3.4cm,angle=-90,]{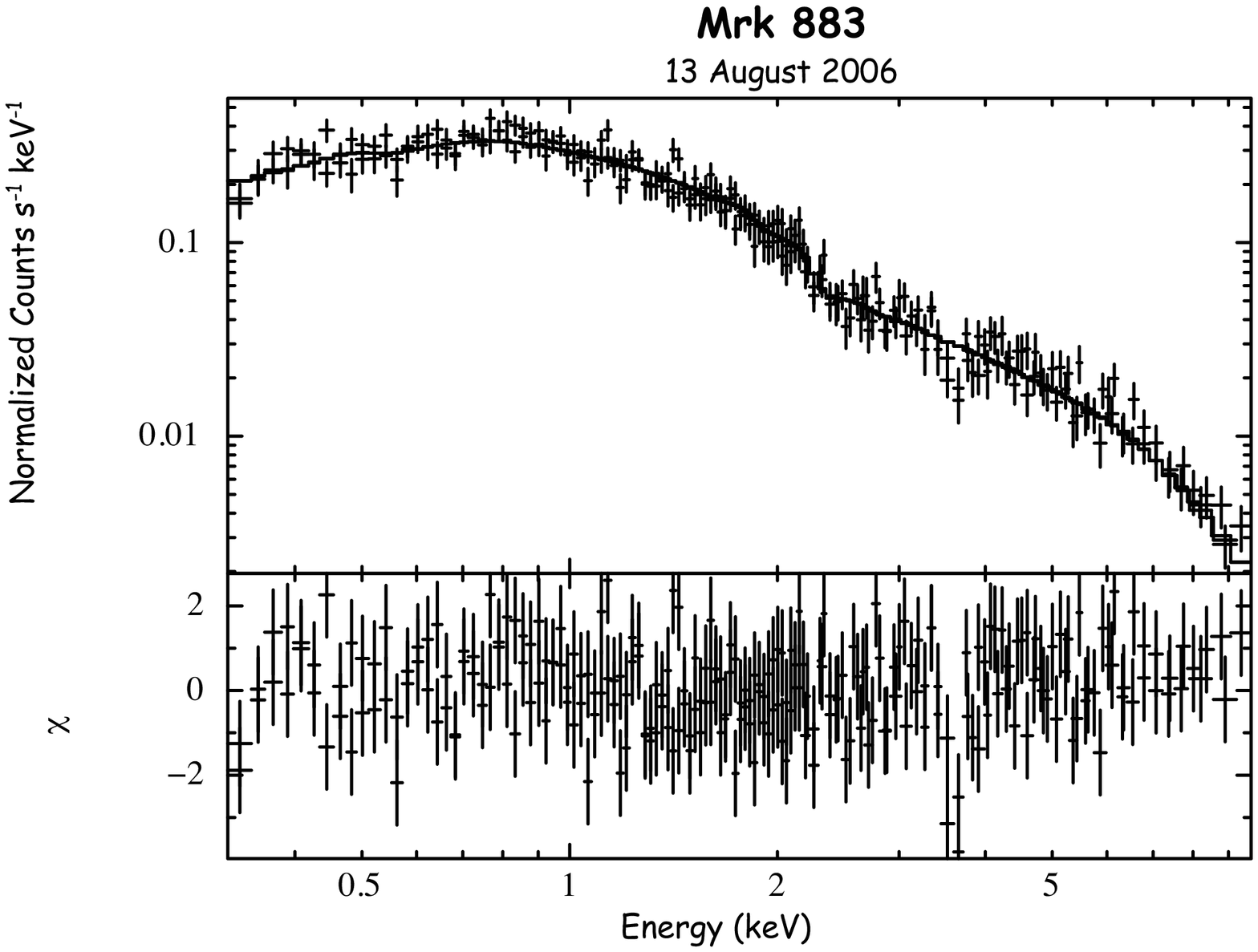} &
\hspace*{\columnsep}\hspace*{\columnsep}
\includegraphics[height=2.9cm,width=3.4cm,angle=-90]{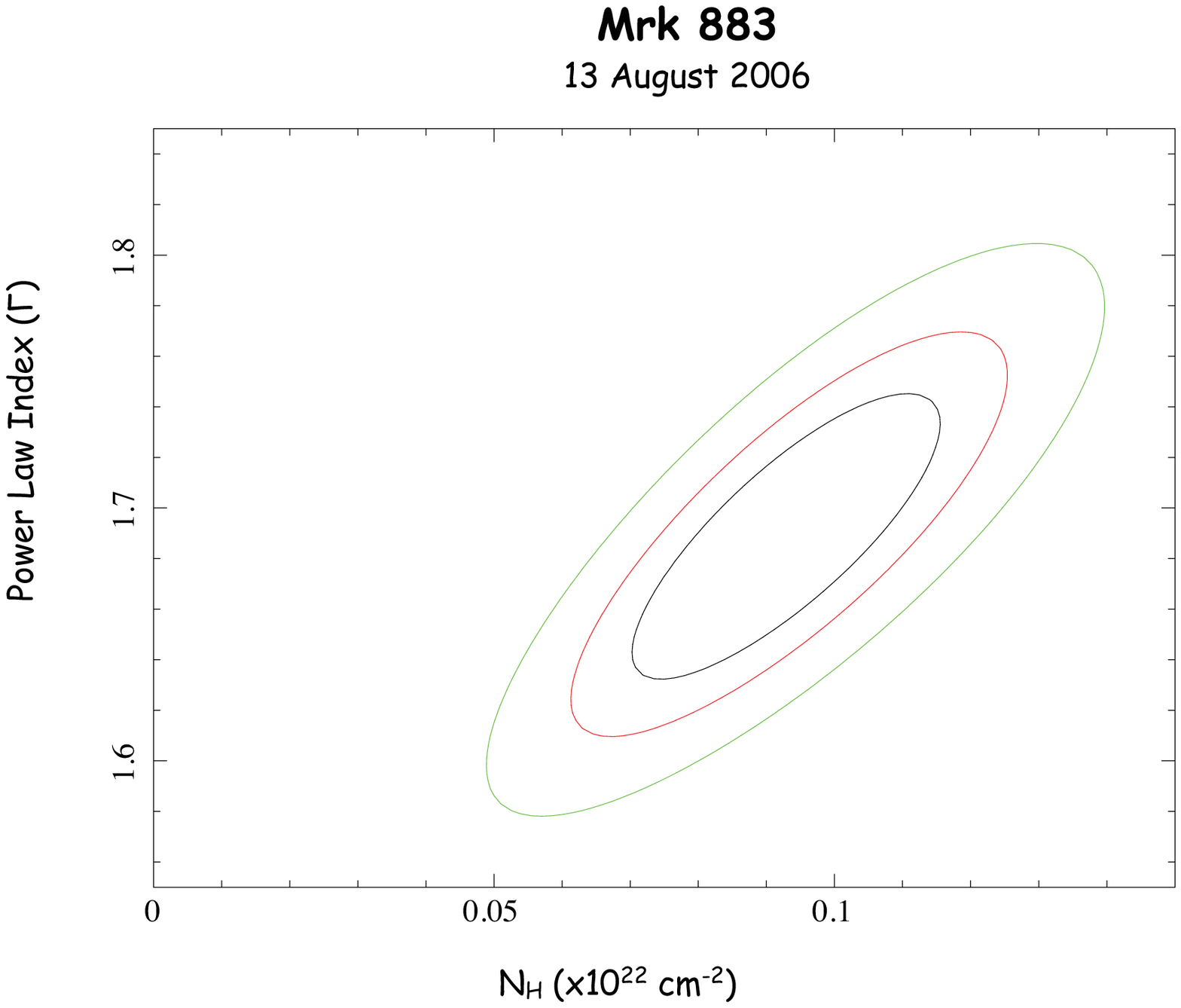} \\
 \includegraphics[height=6.4cm,width=3.4cm,angle=-90]{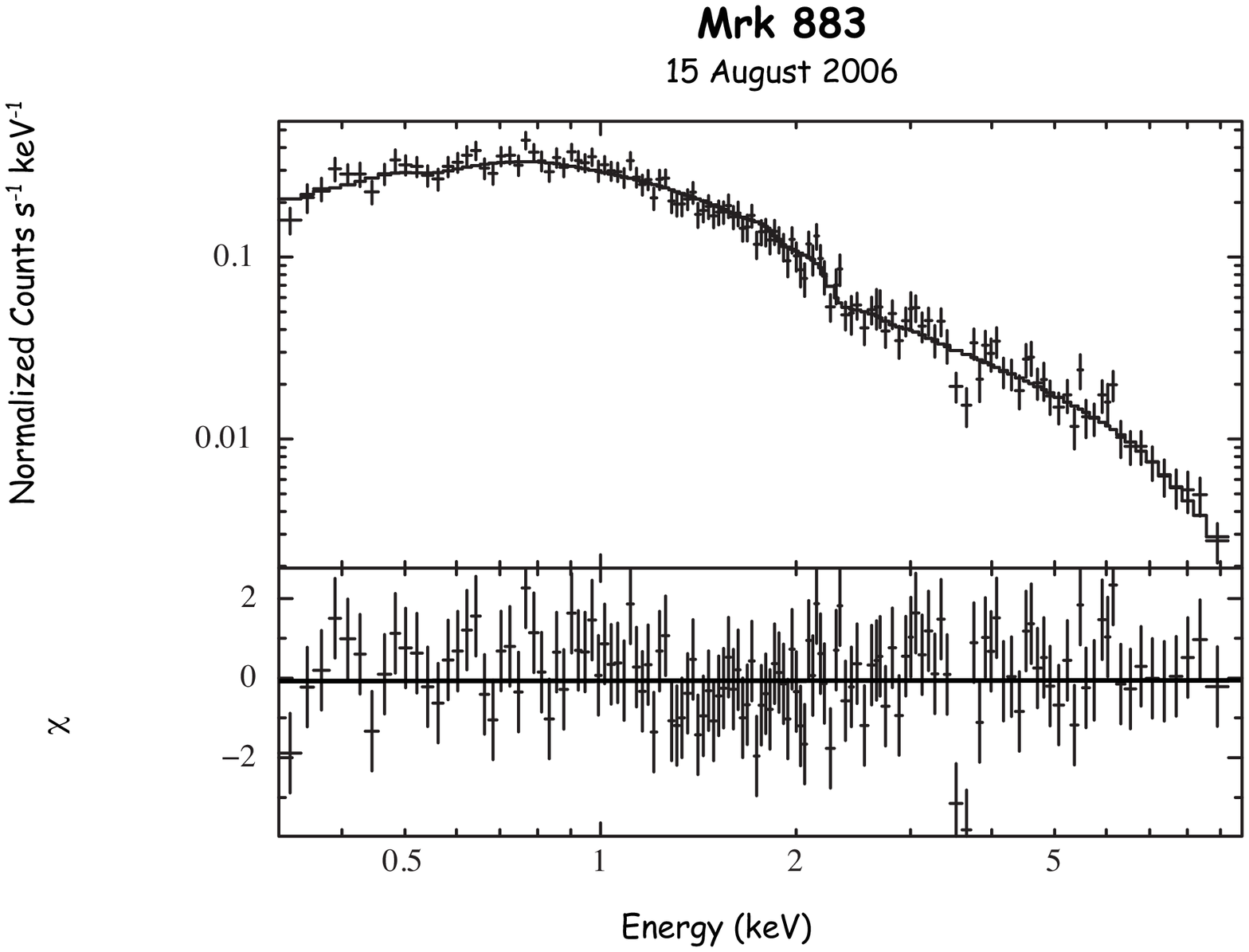} &
\hspace*{\columnsep}\hspace*{\columnsep}
\includegraphics[height=2.9cm,width=3.4cm,angle=-90]{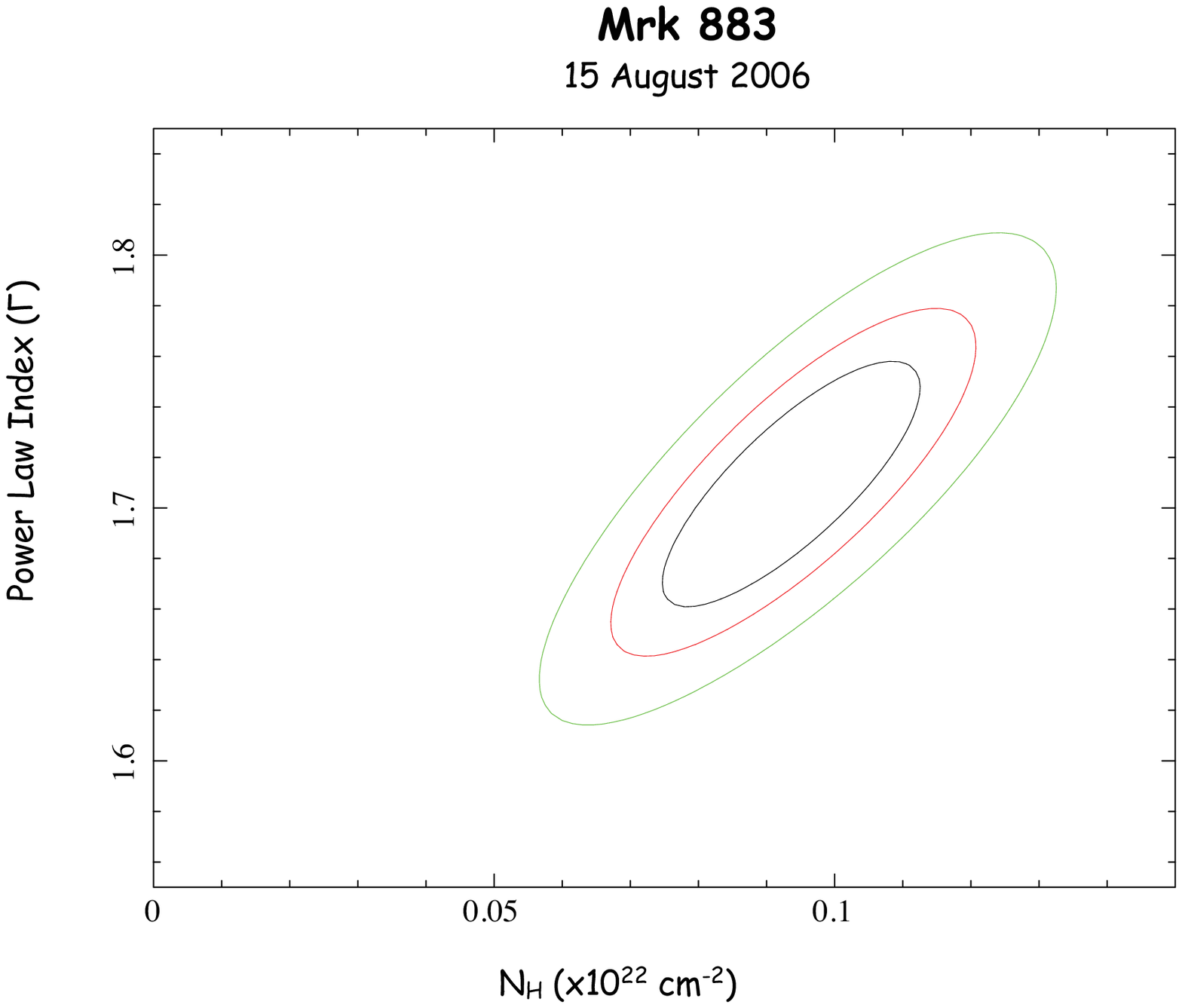} \\
 \includegraphics[height=6.4cm,width=3.4cm,angle=-90]{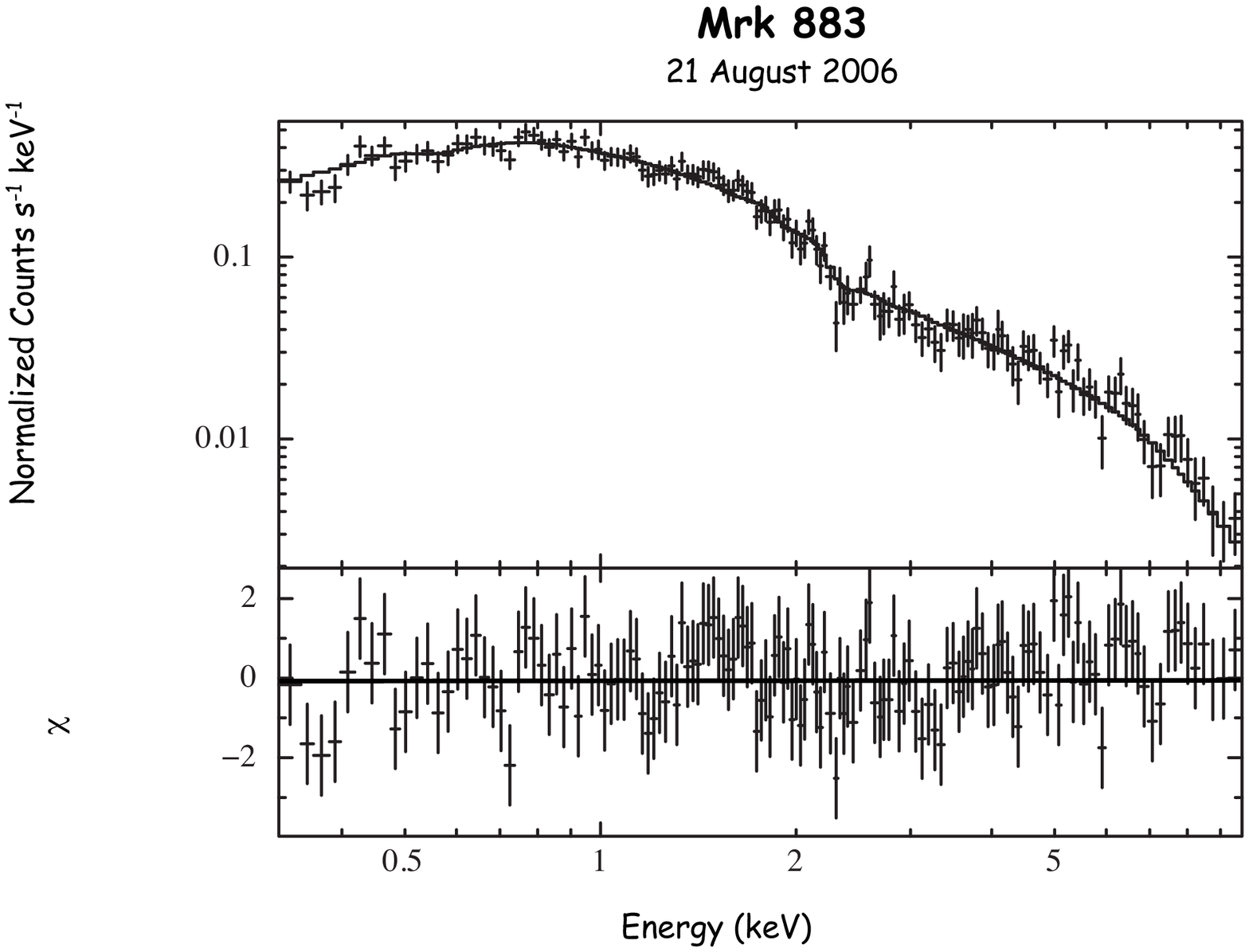} &
\hspace*{\columnsep}\hspace*{\columnsep}
 \includegraphics[height=2.9cm,width=3.4cm,angle=-90]{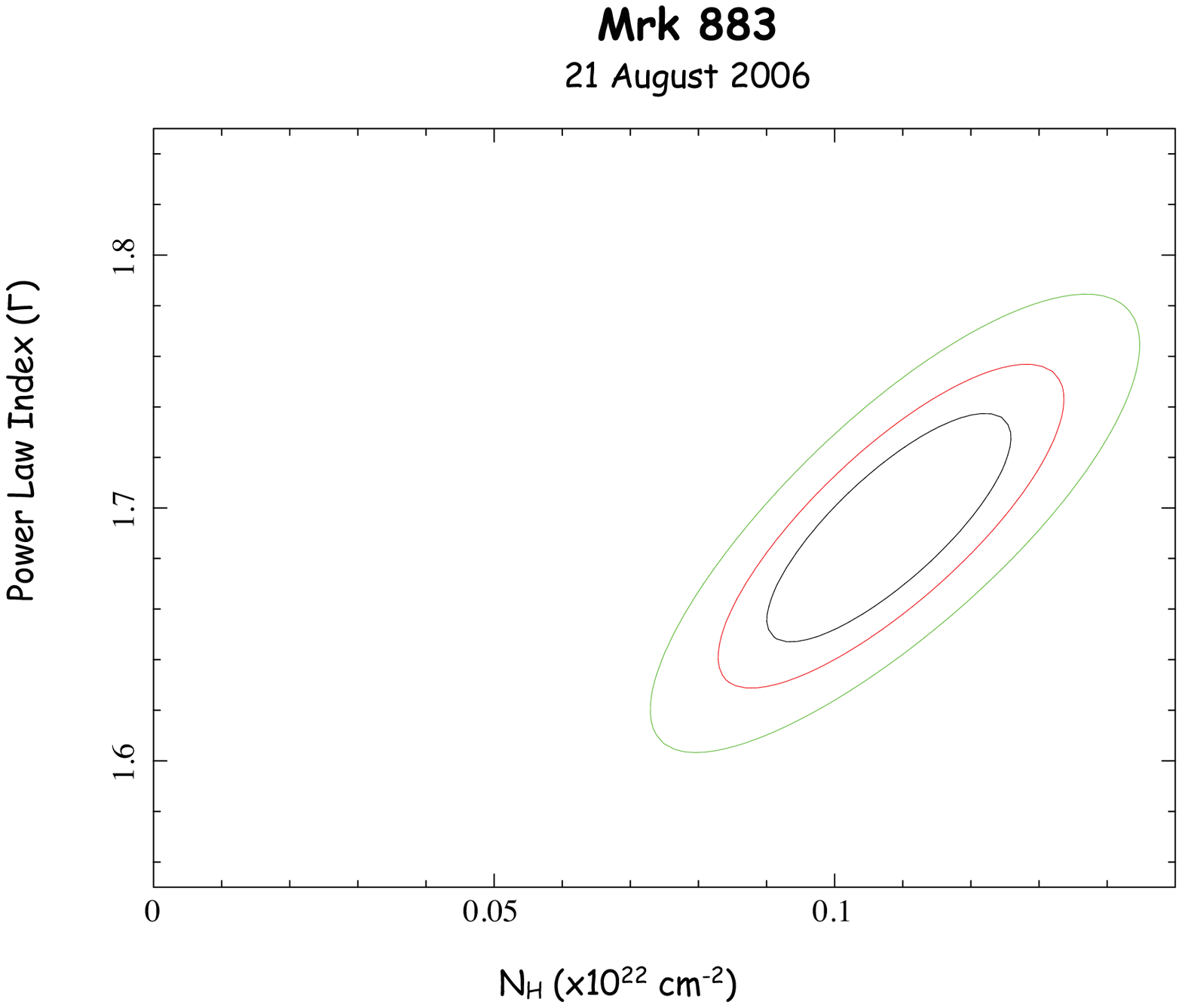}
 \end{tabular}
\end{center}
\caption{Left column: EPIC-pn observed  spectra of SDSS
J162952.88+242638.3 (Mrk~883) together with the best-fit model  and
the residuals. More details of the model can be found in
Table~\ref{tab:xray}. Right column: contour plots of the index of the
power law and the equivalent hydrogen column density for the three
XMM-Newton observations of this object. Each line corresponds to a
different observation date.}
\label{fig:xray}
\end{figure*}

\begin{deluxetable}{cccccc}
\tabletypesize{\scriptsize}
\tablecolumns{6}
\tablewidth{0pc}
\tablecaption{Intermediate-type AGN Sample\label{tab:sample}}
\tablehead{
\colhead{Galaxy} & \colhead{SDSS}  & \colhead{Other} & \colhead{$z$} & \colhead{$D_{L}$}  &
 \colhead{$M_{g}$} \\
\colhead{\#} & \colhead{ID} & \colhead{Name} &  & \colhead{(Mpc)}}
\startdata
1   & \object{J120655.63+501737.1}  & \object{SBS 1204+505B}  & 0.062 &  278.01 & 16.20    \\
2   & \object{J121600.04+124114.3}  & \object{Mrk 764}                & 0.066 &  296.78 & 14.87  \\
3   & \object{J121607.08+504930.0}  & \object{Mrk 1469}              & 0.031 &  135.92 & 15.27   \\
4   & \object{J141238.14+391836.5}  & \object{NGC 5515}            & 0.026 &  113.58 & 13.78   \\
5   & \object{J143031.18+524225.8}  & \object{SBS 1428+529}     & 0.044 &  194.77 & 15.25     \\
6   & \object{J144049.35+505009.2}  & \object{SBS 1439+510}     & 0.117 &  544.65 & 17.28  \\
7   & \object{J153810.05+573613.1}  & \object{SBS 1537+577}     & 0.073 &  329.87 & 15.51    \\
8   & \object{J162952.88+242638.3}  & \object{Mrk 883}                 & 0.038 & 167.47 &  15.26  \\
9   & \object{J212851.19--010412.4}  & \object{IC 1385}                 & 0.052 &  231.51 & 15.21  \\
10 & \object{J234428.81+134946.0}  &  \nodata                             & 0.068 &  306.21 & 16.32 \\ 
\enddata
\tablecomments{Col. 1: galaxy identification. Col. 2: SDSS id. Col. 3: other names. Col. 4: redshift by NED. Col. 5: luminosity distance. Col. 6: 
Petrosian$g$-band magnitude given in SDSS-DR7.}
\end{deluxetable}

\landscape
\begin{deluxetable}{ccccccccccccc}
\tabletypesize{\scriptsize}
\tablecolumns{13}
\tablewidth{0pc}
\tablecaption{Optical Spectroscopic Parameters\label{tab:fits}}
\tablehead{
\colhead{Galaxy} & \colhead{S/N} & \colhead{FWHM (H$\alpha$)} & \colhead{FWHM (H$\beta$)}
& \multicolumn{3}{c}{$\log$ {\textsc{[O\,III]}}/H$\beta$} &  
\multicolumn{3}{c}{$\log$ {\textsc{[N\,II]}}/H$\alpha$} &  \multicolumn{3}{c}{$\log$ {\textsc{[S\,II]}}/H$\alpha$}  \\
\#     &     & \colhead{(km~s$^{-1}$)}    & \colhead{(km~s$^{-1}$)} & \colhead{blue} & \colhead{red}
& \colhead{single-peak} & \colhead{blue} & \colhead{red}         & \colhead{single-peak}
& \colhead{blue} & \colhead{red}     & \colhead{single-peak}
}
\startdata
1  & 13.5$\pm$0.4   & 2950$\pm$58      & 2565$\pm$284               & 0.96$\pm$0.06 & 0.20$\pm$0.01   &                            & -0.31$\pm$0.02   & -0.17$\pm$0.10    &                             & -0.54$\pm$0.03 & -0.57$\pm$0.03  &                          \\
2  &  9.2$\pm$0.2    & 4860$\pm$158    & 5549$\pm$449$\ast$     &                           &                           & 0.18$\pm$0.01   &                             &                               & -0.34$\pm$0.02  &                           &                             & -1.14$\pm$0.05\\
3  &  8.8$\pm$0.2    & 6700$\pm$69      & 6190$\pm$739               & 1.06$\pm$0.04  & 1.24$\pm$0.05  &                            & -0.13$\pm$0.01   & 0.19$\pm$0.01      &                            & -0.41$\pm$0.02  & 0.23$\pm$0.01  &                           \\
4  &  8.3$\pm$0.2    & 3551$\pm$279    & 4068$\pm$489$\ast$     & 0.35$\pm$0.01  & 0.34$\pm$0.01  &                            & -0.06 $\pm$0.002 & -0.07$\pm$0.002  &                            &-0.34$\pm$0.01  & -0.38$\pm$0.01  &                            \\
5  &  5.9$\pm$0.1    & 3878$\pm$341    & 4438$\pm$583$\ast$     & 0.12$\pm$0.01  & 0.37$\pm$0.02  &                            & -0.19$\pm$0.01    & -0.13$\pm$0.01    &                            &-0.53$\pm$0.02  & -0.52$\pm$0.02  &                            \\
6  &  6.5$\pm$0.4    & 3194$\pm$143    & 2584$\pm$519               &                           &                           & 0.32$\pm$0.04   &                              &                              & -0.15$\pm$0.02  &                           &                             &-0.51$\pm$0.06\\
7  & 10.6$\pm$0.3   & 2491$\pm$38      & 2788$\pm$187               &                           &                           & 0.16$\pm$0.01   &                              &                              & -0.34$\pm$0.02  &                           &                             &-0.64$\pm$0.04\\
8  & 11.5$\pm$0.3   & 2776$\pm$87      & 3187$\pm$219$\ast$     & 0.76$\pm$0.04  & 0.18$\pm$0.01  &                             & -0.29$\pm$0.02   & -0.35$\pm$0.02    &                            &-0.12$\pm$0.01  & -0.48$\pm$0.02   &                             \\
9  & 10.7$\pm$0.4   & 1955$\pm$25      & 2443$\pm$101               &                          &                           & 0.42$\pm$0.03    &                              &                              & -0.24$\pm$0.02  &                           &                             &-0.66$\pm$0.05\\
10 &  6.3$\pm$0.5   & 2094$\pm$38      & 2442$\pm$178               &                          &                           & 0.52$\pm$0.10    &                              &                              & -0.36$\pm$0.07  &                           &                             &-0.70$\pm$0.13\\
\enddata
\tablecomments{Col. 1: Galaxy \# (cf. Table~\ref{tab:sample}). Col. 2: SDSS spectra S\,/N obtained from 
STARLIGHT \citep{2005MNRAS.358..363C}. Col. 3: FWHM(H$\alpha$) broad component. Col. 4: FWHM(H$\beta$) broad component. 
Cols. 5-7: $\log$ {\textsc{[O\,III]}}/H$\beta$ line ratios for  double-peaked objects (blue and red) and
for single-peaked sources. Col. 8-10:$\log$ {\textsc{[N\,III]}}/H$\beta$ line ratios for  double-peaked objects (blue and red) and
for single-peaked sources. Col 11-13: $\log$ {\textsc{[S\,II]}}/H$\alpha$ line ratios for double and single-peaked sources. Note that we mark an $\ast$
in objects where the FWHM(H$\beta$) was obtained using \citet{2008AJ....135..928S}, see text. }

\end{deluxetable}

\begin{deluxetable}{cccccc}
\tabletypesize{\scriptsize}
\tablecolumns{6}
\tablewidth{0pc}
\tablecaption{Velocity differences in narrow double-peaked AGN\label{tab:deltav}}
\tablehead{
\colhead{SDSS} & \colhead{J120655.63+501737.1} &
\colhead{J121607.08+504930.0} & \colhead{J141238.14+391836.5} &
\colhead{J143031.18+524225.8}  & \colhead{J162952.88+242638.3} \\ 
& \colhead{(\#1)} &  \colhead{(\#3)} & \colhead{(\#4)} & \colhead{(\#5)} & \colhead{(\#8)} \\
\hline
 & \colhead{$\Delta V$} &  \colhead{$\Delta V$} & \colhead{$\Delta V$}
& \colhead{$\Delta V$} & \colhead{$\Delta V$} \\
 Line     &  \colhead{(km~s$^{-1}$)}  & \colhead{(km~s$^{-1}$)} &
\colhead{(km~s$^{-1}$)} &
\colhead{(km~s$^{-1}$)} &  \colhead{(km~s$^{-1}$)}
}
\startdata
H$\alpha$$\lambda$6562.74                    &  116$\pm$17          & 209$\pm$122    &   264$\pm$15     &   296$\pm$5     & 115$\pm$3    \\
{\textsc{[N\,II]}}$\lambda$6548                  &  117$\pm$31           & 177$\pm$48     &   232$\pm$28    &   276$\pm$32   & 110$\pm$17   \\
{\textsc{[N\,II]}}$\lambda$6583                  &  118$\pm$5             & 160$\pm$107   &   231$\pm$32    &   274$\pm$12   & 135$\pm$5    \\
{\textsc{[S\,II]}}$\lambda$6716                  &  111$\pm$15           & 97$\pm$9          &   226$\pm$7      &   272$\pm$15   & 128$\pm$7    \\
{\textsc{[S\,II]}}$\lambda$6731                  &   96$\pm$35            & 56$\pm$8         &   226$\pm$13    &   271$\pm$20   & 123$\pm$7   \\
H$\beta$$\lambda$4861.29                      &  120$\pm$11           & 160$\pm$10     &   263$\pm$80    &   264$\pm$17   & 142$\pm$20   \\
 {\textsc{[O\,III]}}$\lambda$4959                &  118$\pm$35           &132$\pm$26      &   258$\pm$18    &  199$\pm$185  & 139$\pm$10    \\
{\textsc{[O\,III]}}$\lambda$5007                &  116$\pm$11           &120$\pm$7         &   256$\pm$120  &  198$\pm$64    & 130$\pm$3    \\
\hline
average value  &  114$\pm$7  &  139$\pm$45  & 245$\pm$16  &  256$\pm$34  &  128$\pm$11  \\
\enddata
\tablecomments{Velocity differences obtained between the red and the blue Gaussian components ($\Delta V=V_{red} - V_{blue}$). 
Col. 1: line rest frame. Cols. 2-6: velocity differences obtained for SDSS objects J120655.63+501737.1,
J121607.08+504930.0, J141238.14+391836.5, J143031.18+524225.8, and J162952.88+242638.3, respectively. Last row indicates the average 
$\Delta V$ value and standard deviation of each object considering all lines
}
\end{deluxetable}

\begin{deluxetable}{ccccc}
 \tabletypesize{\scriptsize}
\tablecolumns{5}
\tablewidth{0pc}
\tablecaption{Luminosities and Eddington Rates\label{tab:lumin}}
\tablehead{
\colhead{Galaxy}   & \colhead{$\lambda\,L_{\lambda5100}$} & \colhead{$L_{bol}$} & \colhead{$L_{Edd}$} & 
\colhead{$L_{bol}/L_{Edd}$} \\
\colhead{\#} & \colhead{(10$^{41}$ erg~s$^{-1}$)} & \colhead{(10$^{42}$ erg~s$^{-1}$)} &
\colhead{(10$^{44}$ erg~s$^{-1}$)} & \colhead {(10$^{-2}$)}
}
\startdata
1    & 49.92$\pm$6.98   & 45$\pm$6.28       & 15.05$\pm$5.08     & 2.98$\pm$1.42   \\
2    & 14.52$\pm$4.84   & 13.07$\pm$4.36  &  37.99$\pm$14.24  & 3.44$\pm$0.24   \\
3    & 42.02$\pm$3.65   & 37.82$\pm$3.28  &  80.45$\pm$26.41  & 0.47$\pm$0.20   \\
4    & 5.77$\pm$1.53     & 5.19$\pm$1.38    & 12.87$\pm$5.39     & 0.40$\pm$0.28   \\
5    &  2.11$\pm$0.41    & 1.90$\pm$0.37    & 9.27$\pm$5.38       & 0.21$\pm$0.16   \\
6    &  40.43$\pm$6.83  & 36.39$\pm$6.14  & 13.77$\pm$7.32     & 2.64$\pm$0.02   \\
7    &  45.47$\pm$4.90  & 40.92$\pm$4.37  & 16.98$\pm$3.96     & 2.41$\pm$0.82   \\
8    &  53.74$\pm$7.17  & 48.37$\pm$6.45  &  24.12$\pm$6.03    & 2.01$\pm$0.01   \\
9    &  18.14$\pm$1.63  & 16.33$\pm$1.47  &   8.23$\pm$1.43     &1.98$\pm$0.52    \\ 
10  &   5.08$\pm$1.76   & 4.57$\pm$1.59    &   4.35$\pm$1.59     &1.11$\pm$0.75   \\

\enddata

\tablecomments{ Col. 1: galaxy \# (cf. Table~\ref{tab:sample}). Col. 2:  luminosity at 5100 \AA\, measured using a continuum power-law fit. Col. 3: bolometric luminosities obtained assuming $L_{bol}\,=\,9\lambda L_{\lambda5100}$. Col. 4: Eddington luminosities using our estimated M${_BH}$. Col. 5: Eddington ratio.
}
\end{deluxetable}

\begin{deluxetable}{cccccc}
\tabletypesize{\scriptsize}
\tablecolumns{6}
\tablewidth{0pc}
\tablecaption{BH Mass\label{tab:BHm}}
\tablehead{
\colhead{Galaxy}& \colhead{$\sigma_{\star}$}  & \colhead{$\log M_{BH}$} & \colhead{Spectral Classification}& \colhead{Double-peak} &
\\
\colhead{\#} & \colhead{(km~s$^{-1}$)}  & \colhead{(M$_{\odot})$} 
}
\startdata
1   &   90$\pm$6   & 7.08$\pm$ 0.14 & Sy 1.8\,+\,SB/Sy 1.5\,+\,SB & yes \\
2   & 111$\pm$6   & 7.48$\pm$0.16 & Sy 1.9\,+\,SB/SB &  no\\
3   & 147$\pm$7   & 7.81$\pm$ 0.14 & Sy 1.8 & yes$^{\star}$ \\
4   & 195$\pm$8   & 7.01$\pm$ 0.18 & Sy 1.9 & yes$^{\star}$ \\
5   & 165$\pm$9   & 6.86$\pm$0.25 & Sy 1.5\,+\,SB & yes\\
6   & 188$\pm$17 & 7.04$\pm$0.23 & Sy 1.5\,+\,SB & no \\
7   & 91$\pm$6     & 7.13$\pm$0.10 & Sy 1.5\,+\,SB & no\\
8   & 187$\pm$9   & 7.28$\pm$0.11 & Sy 1.9\,+\,SB & yes\\
9   & 79$\pm$6     & 6.82$\pm$0.08 & Sy 1.5\,+\,SB & no\\
10 & 93$\pm$11   & 6.54$\pm$0.16 & Sy 1.8\,+\,SB & no\\
\enddata

\tablecomments{Col. 1: galaxy id. (cf. Table~\ref{tab:sample}). Col. 2: velocity dispersion obtained with STARLIGHT. Errors were estimated from
our Monte Carlo simulation on the data, see text. Col. 3: Black hole mass estimates were obtained with the relation given by 
\citet{2006ApJ...641..689V}. Black hole mass for object \#8 have to  be taken with care since the host galaxy is probably a merger. Col. 4: Sy type. 
Col. 5: yes if it is double peak, and $^{\star}$ shows the two dual AGN candidates found in our study.}
\end{deluxetable}

\landscape
\begin{deluxetable}{ccccccccccccc}
\tabletypesize{\scriptsize}
\tablecolumns{13}
\tablewidth{0pt}
\tablecaption{X-ray Spectroscopic Parameters\label{tab:xray}}
\tablehead{
\colhead{\bf Obs.} & & \colhead{\bf Neutral Abs}   & \colhead{\bf Power Law} & \multicolumn{2}{c} {\bf Emission Line} & \multicolumn{3}{c} {\bf Goodness}& \multicolumn{2}{c} {\bf Flux}  & \multicolumn{2}{c} {\bf Luminosity}
} 
\startdata
Day  &  Model & N$_H$ & $\Gamma$  & Energy & EW & $\chi^2_{\nu}$ & d.o.f. & F-test & 0.5-2 keV & 2-10 keV & 0.5-2 keV & 2-10 keV \\
\hline\\
13 & A & 9$\pm$2 & 1.68$\pm$0.06 &  &  & 134.0 & 118 &   & $0.65^{+0.03}_{-0.04}$ & $1.51\pm0.05$ & $2.64^{+0.12}_{-0.16}$ & $4.90\pm0.16$ \\
     & B & 9$\pm$2 & 1.69$\pm$0.06 & 6.4f & $<$170 & 133.9 & 117 & 35.2\% \\
     & C &10$\pm$2 & 1.70$\pm$0.06 & 6.88$^{+0.08}_{-0.15}$ & 210$^{+180}_{-160}$ & 129.0 & 116 & 87.5\% \\

15 & A &  9$\pm$2  &  1.71$\pm$0.05 &   &  & 135.0 & 129 &  &  0.66$\pm$0.02 & 1.47$\pm$0.05 & 2.66$\pm$0.08 & 4.79$\pm$0.16  \\
     & B & 10$\pm$2 &  1.72$\pm$0.05 & 6.4f   & 210$^{+90}_{-150}$ & 128.9 & 128 & 98.5\% \\
     & C & 10$\pm$2 &  1.72$\pm$0.05 & 6.34$^{+0.08}_{-0.10}$ & 200$^{+130}_{-120}$ & 127.3 & 127 & 97.6\% \\

21 &A& 11$\pm$2 & 1.69$\pm$0.05 &   &  & 128.9 & 145 &  & 0.83$\pm$0.03 & 1.95$\pm$0.05 & 3.43$\pm$0.12 & 6.33$\pm$0.16 \\
     &B& 11$\pm$2 & 1.70$\pm$0.05 &  6.4f & 75$^{+4}_{-3}$ & 126.9 & 144 & 86.6\% \\
     &C& 11$\pm$2 & 1.70$\pm$0.05 &  6.55$\pm$0.15 & 150$\pm$110 & 123.9 & 143 & 94.1\% \\

\enddata

\tablecomments{Col 1: day of observations, data are from August 2006. Col 2: model A: absorbed  power  law; model  B: absorbed power law  plus a neutral iron line with the energy fixed; model C: absorbed power law 
plus an iron line with the energy left free to vary. Col 3: neutral absorption in units of 10$^{20}$\,cm$^{-2}$. Co (4): power law index. Col (5) and (6): energy in keV and equivalent width in eV, respectively. Col (7): 
modified reduced $\chi^{2}$. Col (8)  degrees of freedom. Col (9) F-test. Col. 10 and 11: fluxes in 10$^{-12}$\,erg\,s$^{-1}$\,cm$^{-2}$. Col. 12 and 13: luminosities in 10$^{42}$erg\,s$^{-1}$.
}

\end{deluxetable}

\end{document}